\documentclass[preprint,aps,amssymb]{revtex4}

\usepackage{graphicx}
\usepackage{epsfig}
\usepackage{longtable}
\usepackage{longtable}

\begin{document}

\title{Candidates for three-quasiparticle $K$-isomers in odd-even Md  -  Rg nuclei}

\author{P.~Jachimowicz$^{1}$}
\author{M.~Kowal$^{2}$} \email{michal.kowal@ncbj.gov.pl}
\author{J.~Skalski$^{2}$}

\affiliation{$^1$ Institute of Physics, University of Zielona G\'{o}ra, Z. Szafrana 4a, 65-516 Zielona G\'{o}ra, Poland}
\affiliation{$^2$ National Centre for Nuclear Research, Pasteura 7, 02-093 Warsaw, Poland}

\date{\today}

\begin{abstract}

 We performed a search for three-quasiparticle high-$K$ isomer candidates  in odd-even Md - Rg nuclei by considering the
lowest lying 1$\pi$2$\nu$ and 3$\pi$ excitations. Our approach involves calculating the energies of different nuclear
configurations using a microscopic-macroscopic model with the Woods-Saxon potential. We explore three pairing scenarios:
blocking, quasi-particle BCS method, and particle number projection formalism. The optimal deformations for both ground-states
and high-$K$ configurations are determined through a four-dimensional energy minimization process. By analyzing the obtained
excitation energies, we discuss the most promising candidates for high-$K$ isomers and compare them, where possible,
with existing experimental data. We also discuss a possible isomer $\alpha$-decay hindrance by using calculated $Q_{\alpha}$-hindrances.

% By selecting the lowest lying  1$\pi$2$\nu$ and 3$\pi$ excitations
% we found the candidates for $K$-isomers in odd-even Md - Rg nuclei.
%  Energies of nuclear configurations are calculated within the
% microscopic-macroscopic model with the Woods-Saxon potential in three
% pairing scenarios: via blocking, using quasi-particle BCS method and
% in the particle number projection formalism.
% Optimal deformations for ground-states and high-$K$ configurations
% are found by the four-dimensional energy minimization.
% Obtained excitation energies are used to discuss the most favoured
% candidates for high-$K$ isomers and, as far as possible, to compare
% them with available experimental data.

\end{abstract}

%\pacs{21.10.-k, 21.60.-n, 27.90.+b}

\maketitle

\section{INTRODUCTION}

Isomeric states in the domain of superheavy (SH) nuclei are of a considerable interest. Finding them experimentally
not only provides clues to a scheme of low-lying single-particle (s.p.) orbits at large $Z$ and $N$ which could be
checked against theoretical models but also offers a chance of
coming upon longer-lived states of SH isotopes which would then provide new opportunities for their study - see \cite{Walker1999,Batar,Lopez,Herzberg}.
In the present work, we give candidates for the three-quasiparticle (3-q.p.) high-$K$
\footnote {The value in question, typically denoted as a $K$ quantum number by the Nilsson notations \cite{Nilsson}, represents the
sum of the total spin projections of the individual single-particle orbitals involved.}
isomers in odd-even Md - Rg nuclei, which follow from the Woods-Saxon microscopic-macroscopic (MM) model extensively studied
in the region of heavy and SH nuclei.

The enhanced stability, or half-life, of some multi-quasiparticle states with high angular momentum and $K$ quantum numbers, situated at relatively low  excitation energies, has been investigated a lot in the medium-mass and deformed nuclei \cite{Walker2016,kon2}. It  predominantly results from the retardation of electromagnetic transitions -  gamma-rays or internal conversion electrons -
\footnote{fission probabilities for such nuclei are exceedingly low not only at their ground states but also at their excited states,
even up to several MeV, where the majority of high-$K$ states are found} connecting states with a sizable difference in $K$ and the absence of typical de-excitation modes  with $\Delta K\approx 0$. Both circumstances come into play for
 a particular placement of a high-$K$ configuration in the nuclear level scheme, when all states below it have
 significanly smaller $K$.

A recent progress in experimental studies on excited states in
even-even superheavy nuclei, including isomeric ones, allowed for some checks of theoretical models utilized in this region.
While the calculated and measured excitation
energies can be easily compared, the assignment of particular  configurations to the discovered isomeric states often poses substantial difficulties. Typically, the assignments
suggested in the literature are  based on theoretical models, a knowledge of states in neighboring nuclei, and the systematic patterns observed.

Already 50 years ago, the initial observation of the 0.28 s isomer in $^{254}$No was conducted \cite{Ghiorso1973}; its isomeric character was confirmed and attributed to the two-quasiproton $K^{\pi}=8^{-}$ configuration in \cite{Tandel2006}.
Its measured energy, 1293 \cite{Herzberg2006,HerzbergNAT2006}, 1296 \cite{Tandel2006}, 1295(2) \cite{Hessberger2010},
and 1297(2) \cite{Clark2010} keV can be compared to values in the range $\langle 1.1 \div 1.5 \rangle$~MeV, obtained in various  MM approaches (see Table III in \cite{Theisen2015}).
 The experimentally measured  excitation energy of the $T_{1/2}
 = 109(6)~ms$ isomer in $^{252}$No is approximately 1.25~MeV, as reported in \cite{Robinson2008,Sulignano2012}.
Recently, Kallunkathariyil et al. investigated the stability of the 35 $\mu$s isomer in the neighboring $^{250}$No \cite{Kallunkathariyil2020}.
In the previous study by Peterson et al. \cite{Peterson2006}, its tentative assignment was suggested as
 two-quasineutron  $K^{\pi} = 6^{+}$,
${\nu5/2^{+}[622] \otimes \nu 7/2^{+}[624]}$ configuration. However, its experimental energy is currently unknown.
 Finally, in a quite recent measurement on the Fragment Mass Analyzer at ANL, two isomers, 247 and 4.7 $\mu$s, were identified in  $^{254}$Rf.
 %%data were obtained using digital electronics under vacuum %conditions at the
  David et al. suggested that the shorter-lived one corresponds to a two-quasiparticle configuration, either
 $\nu^2 8^{-} = \{{\nu 9/2^{-}[734] \otimes \nu 7/2^{+}[624]}\}$ or $\pi^2 8^{-}=\{\pi7/2^{-}[514] \otimes \pi9/2^{+}[624]\}$, while the other one to the
 4-quasiparticle configuration ${\pi^2 8^{-} \otimes \nu^2 8^{-} }$ \cite{David2015}.

Concerning experimental studies on odd-even nuclei, a 1.4(1) ms isomer in $^{255}$Lr was  discovered by using  tunnel detectors of the GABRIELA
facility \cite{Hauschild2008}.
By analyzing the coincidences between isomeric conversion electrons (ICEs) and gamma rays, the  lower limit of 720 keV was established for the
isomer excitation energy. Subsequently, this isomer was also observed at GSI and Berkeley Labs \cite{GSI,Berkeley}.
%In these experiments,
%the decay of the isomer was detected by observing the cascade of conversion electrons and coincident atomic radiation at the same position as
%the implanted recoil in the focal plane Si-strip detector.
Recently, two isomers, 2.8 ms at an excitation energy $\geq$ 910 keV in
$^{249}$Md, and 1.4 ms at $\geq$ 844 keV in $^{251}$Md were reported in \cite{Goigoux2021}, following the previous work on spectroscopy in both isotopes
\cite{Chatillon,Asai2015,Theissen2020}. As discussed in \cite{Lopez2}, the configuration assignments for these Lr and Md isomers may be still considered controversial and related to
the nature of the $K^{\pi}= 8^-$ state in $^{254}$No. While its two-neutron configuration is favoured by most interpretations,
%Theoretical models suggest that the energy separation between the $\pi 9/2^+[624]$ and $\pi 7/2^-[514]$ states should be
%pretty large, which is at odds with the conclusions of Jeppesen et al. \cite{Berkeley}.  is still under debate, and
only a measurement of the splitting of its hyperfine structure can determine whether the state is
based on a two-quasiproton or two-quasineutron excitation or a mixture of both (in spite of the very different intrinsic $g_K$ factors: $g_K\approx 1$ for the  $\pi 7/2^-[514]$ and $\pi 9/2^+[624]$ 2q.p. state, and $g_K \approx -0.28$ for $\nu 7/2^+[613]$ and $\nu 9/2^-[734]$, $|g_K - g_R|$ controlling the intra-band decays should be similar, with the rotational g factor $0.7 Z/A \leq g_R \leq Z/A$ -  see discussion in \cite{Lopez2}).
The analysis of experimental $\alpha$-decay data of $^{257}$Db and its daughter products led to the conclusion that isomeric states undergoing $\alpha$-decay exist in $^{257}$Db and $^{253}$Lr. \cite{Hessberger2001}.
%Additionally, two previously unknown isotopes, %$^{256}$Db and $^{252}$Lr,
%were synthesized using the highest bombarding energies of 4.97 AMeV and 5.08 AMeV, respectively.

The most up-to-date information regarding isomers in the heaviest nuclei can be found in the following references:
Ackermann et al. (2015), Asai et al. (2015), Theisen et al. (2015), Dracoulis et al. (2016), Walker et al. (2020), and A. Lopez-Martens with K.
Hauschild (2022) \cite{Ackermann2015,Asai2015,Theisen2015,Dracoulis2016,Walker2020,Lopez2}.

A precise prediction of the high-$K$ isomer would require reliable estimates not only of energies
but also of $EM$ transition probabilities among various nuclear states and that is beyond the reach of the present theory of heavy nuclei. Instead, one
usually finds high-$K$ ``optimal'' (i.e., obtained by the tilted Fermi surface method) configurations (or close to them) and selects those with low enough energies.
For a theoretical overview based on the Nilsson-Strutinsky approach, see the work by Walker et al. (2016) \cite{Walker2016}.
Both experimental studies and theoretical predictions agree that $K$-isomers occur in nuclei near nobelium, rutherfordium and heavier.
The likelihood of $K$-isomer existence can be attributed to the proximity of
 high-$\Omega$ orbitals to the Fermi level and the predicted deformed subshell
 gaps around $Z=100$ and $N=152$ \cite{Greenlees2008}.

In the present work we studied nuclei in the following range of neutron numbers: $N=$142-166 for Md, Lr, Db, $N=$144-166 for Bh, $N=$146-166 for Mt and $N=$148-166 for Rg. Compared to studies of $K$-isomers in even-even nuclei like those in \cite{Liu2014}, the present one (odd-even systems) has to face a significantly  larger number of potential candidates. In order to appreciate the uncertainty of the strength of pairing correlations, we study three pairing versions giving different excitation energies of 3q.p. configurations. We also add some considerations regarding a possible decay of the candidate for isomer to the rotational band built on the one-proton configuration that is included in it. A brief discussion of a possible hindrance
 of the $\alpha$-decay of high-$K$ configurations is included and illustrated in the extreme case of the predicted $Q_{\alpha}$-hindrance.

%When selecting promising candidates some allowance is made for uncertainty in the order of the
%calculated proton levels which seems different from the experimental one in lighter isotopes of Md and Lr.

The calculations and  selection of candidates for high-$K$ isomers are described in Sect. II, results are presented and discussed
in Sect. III and conclusions are given in Sect. IV. Tables with calculated characteristics of the lowest high-$K$ configurations are provided in the supplementary material.

\section{THE METHOD}

Realistic MM or mean-field models predict that Md - Rg nuclei are well-deformed in their ground states with axially- and reflection-symmetric
shapes. This is consistent with the experimentally established characteristics of the rotational bands in the No - Rf region \cite{Herzberg2001,Eeckhaudt2005,Ketelhut2009,Greenlees2012}.
Similar shapes are predicted for excited few-q.p.  configurations (as long as no time-reversal-breaking components are included in the mean field),
and again, the discovered 2-q.p. $K$-isomers in Fm - Rf nuclei with prominent reduced hindrance factors  support this hypothesis. Therefore we assume that
the intrinsic parity of considered states is well defined as is their $K$ - quantum number.

%For the isomeric states in $^{254}$Rf and $^{270}$Ds, the spin-parity assignments predicted in Ref. \cite{Liu2014} are ambiguous, either 8$^{-}$ or  9$^{-}$, due to the limited availability of experimental data. To facilitate a comparative analysis of their decay modes, the same excitation energies of 1 MeV and decay paths for electromagnetic and $\alpha$ decays were assumed. The fission half-lives of approximately $1 \times 10^{-1}$ s and $1 \times 10^{2}$ s were estimated for the 8$^{-}$ and 9$^{-}$ states, respectively \cite{Batar2022}.
%Assuming that the isomer-depopulating  $\gamma$-transition feeds the $I^{\pi}=8^{+}$ member of the ground-state ($K=0$) rotational band  at  excitation energy of 0.5 MeV, one expects its energy of 0.5 MeV, the E1 multipolarity and $\Delta K = 8$ and 9   in $^{254}$Rf and $^{270}$Ds, respectively. Utilizing the most recent parameterization of
%  hindrance factors as a function of $\Delta K$ provided in Ref. \cite{kon2}, half-lives of $7 \times 10^{-3}$ s and $8 \times 10^{-2}$ s were obtained for $^{254}$Rf and $^{270}$Ds, respectively.

% , except time-reversal breaking effects which can break axial symmetry in some degree..
In order to obtain ground-states and configuration-constrained minima, we use a four-dimensional space of deformations $\beta_{\lambda 0}$ defining
the nuclear surface:
\begin{equation}
\label{radius}
R(\theta)= c (\beta) R_0 \left[1+\sum_{\lambda=2,4,6,8}\beta_{\lambda 0} {\rm Y}_{\lambda 0}(\theta)\right],
\end{equation}
where ${\rm Y}_{\lambda 0}(\theta)$ are spherical harmonics, $c(\beta)$ is the volume-fixing factor depending on deformation, and $R_0$ is the
radius of a spherical nucleus.

%calculated as $R_0=1.16 \cdot A^{1/3} fm$.

The MM method we use employs
%correction \cite{Strutinski1966_67} method on the single particle levels obtained after
the deformed Woods-Saxon (WS) potential \cite{Cwiok1987}
%The $n_{p}=450$ lowest proton levels and $n_{n}=550$ lowest neutron levels from the $N_{max}=19$ lowest
%shells of the oscillator are taken into account in the diagonalization procedure.
%Standard values of $\hbar\omega_{0}=41/A^{1/3}~MeV$  for the oscillator energy and
%$\gamma=1.2 \hbar\omega_{0}$ for the Strutinski smearing parameter
%$\gamma$, and a six-order correction polynomial are used in the
%calculation of the shell correction.
and the macroscopic Yukawa-plus-exponential energy model \cite{Krappe1979} with parameters specified in \cite{Muntian2001}. In particular, the used pairing strengths are: $ G_p = (g_{0p} + g_{1p}I)/A;  G_n = (g_{0n} + g_{1n}I)/A; g_{0p} = 13.40 \text{ MeV} ; g_{1p} = 44.89 \text{ MeV} ; g_{0n} = 17.67 \text{ MeV} ; g_{1n} = -13.11 \text{ MeV}$.

%in which deformation dependent Coulomb and surface energies were integrated by using a 64-point Gaussian quadrature.
Parameters of the MM model are kept the same as in all recent applications to heavy and superheavy nuclei which concerned
masses and deformations \cite{Kowal2010}, $Q_\alpha$ - energies \cite{Jachimowicz2014}, the first and second fission barriers in actinides \cite{Jachimowicz2012-20} and SH nuclei \cite{Jachimowicz2017-2}, \cite{Jachimowicz2021}. In the context of many-q.p. excitations, the important feature of the present model is the distinct
subshell gap in the neutron spectrum at $N=152$ around fermium and nobelium, which seems necessary for realistic predictions of $K$-isomeric
states. Two other subshell gaps predicted by the model: at $N=162$ for neutrons and at $Z=108$ for protons, not quite confirmed experimentally yet,
strongly influence the predictions presented here. The very similar Woods-Saxon model was used in \cite{Stefan,ParSob} which can be consulted for
predictions concerning 1-q.p. proton excitations in this  region of nuclei.

The MM method used in this study enables the examination of deformation parameters of higher orders.
In their work \cite{Patyk19911,Patyk19912}, Patyk and Sobiczewski observed a wider shell gap around $Z=100$ and $N=150$ when incorporating $\beta_{60}$ in their definition of the nuclear radius. This modification resulted in an improved agreement with existing experimental data.
Recently, Liu et al. discussed the influence of the deformation parameter $\beta_{60}$ on the properties of high-$K$ isomers in superheavy
nuclei \cite{Liu2011}. The shape parameterization used in the present paper includes still one additional parameter, namely $\beta_{80}$.

%\cite{Kowal2010,Kowal2010_2,Kowal2012,Jachimowicz2012_20,Jachimowicz2013,Jachimowicz2014,archKowal2012,Jachimowicz2017,Jachimowicz2017_2,Jachimowicz2018}

Ground-state- and excited configuration energies are found by the four-dimensional energy minimization
over $\beta_{20}, \beta_{40}, \beta_{60}, \beta_{80}$ (\ref{radius}) performed using the gradient method. To avoid secondary or very
deformed minima the minimization is repeated at least 10 times for each configuration with different starting values of deformations.

The pairing is accounted for by using three procedures: 1) the blocking method in which, after removing selected singly occupied states from the set of doubly occupied orbitals, the BCS energy is calculated on the remaining ones at each step of the procedure of minimization over deformations, 2) a more straightforward quasi-particle method in which
the BCS q.p. energies for blocked nucleons are added to the energy of the even-even core, and the sum is subjected to the minimization over deformation procedure, and 3) the particle-number-projection method (PNP) in which the energy of the particle-number-projected BCS configuration
is minimized over deformations. A short description of the latter procedure is provided in the Appendix.
The reason for including various pairing calculations is a deficiency of the BCS method with blocking, used in our mass model for ground state
(g.s.) properties of odd-$A$ and odd-odd nuclei when applied to many-q.p. excitations. Usually, the blocking method
underestimates excitation energies and pair correlations in many-q.p. states if one uses the pairing strength adjusted to the ground states.

In 2), the microscopic part of the energy for a 1$\pi$2$\nu$ configuration was taken as the sum of BCS
quasi-particle energies of singly occupied levels:
\begin{eqnarray}
\label{quasi}
E_{q.p.}^{*} &=& \sqrt{(\epsilon_{\pi}-\lambda_{\pi})^2+\Delta_{\pi}^2}
\nonumber \\
                       &+&\sqrt{(\epsilon_{\nu_1}-\lambda_{\nu})^2+\Delta_{\nu}^2}   \nonumber
                       +\sqrt{(\epsilon_{\nu_2}-\lambda_{\nu})^2+\Delta_{\nu}^2}     \nonumber  \\
\end{eqnarray}
and the core energy term consisting of the shell- and pairing corrections calculated without blocking. For protons,
the core term, as well as the pairing gap $\Delta_{\pi}$ and the Fermi energy $\lambda_{\pi}$, are calculated for the odd number of particles,
but with the double occupation of all levels. This prescription was used before in \cite{ParSob}.
It gives results similar to those obtained when calculating the shell and pairing correction for the even system with one particle less.
For 3$\pi$ 3-q.p. excitations, the microscopic energy was the sum of three proton-quasiparticle energies and the shell
and pairing corrections calculated without blocking; as for the 1$\pi$2$\nu$ excitations, the odd particle number and double
occupation of levels were used in the BCS procedure for protons.
In the quasiparicle method we use the same pairing strengths as in our mass model.
One can mention that the quasiparticle method underestimates 3-q.p. excitation
energies at particle numbers for which BCS energy gap vanishes or is very small. Admittedly, it is cruder for 3$\pi$ than
for 1$\pi$2$\nu$ configurations due to the larger number of blocked quasiprotons.

 When using the method 3) we face the necessity of adjusting the pairing
 strengths for neutrons and protons to the new PNP procedure.
 Within the BCS method, such adjustment can be performed using experimental
  masses or moments of inertia, like, for example, in \cite{Minkov2022}.
 However, with PNP this becomes quite cumbersome. Therefore,
  we fixed the new strengths $G_n(N,Z)$ for neutrons by so fixing
  the ratio $G_n(N,Z)/G_n^{mod}(N,Z)$, with $G_n^{mod}(N,Z)$ the strength from
  our MM model, as to obtain the energy of the 2-q.p. excitations involving
 two nearly degenerate s.p. levels: the last occupied and the first empty one,
 close to $2\Delta_n({\rm BCS})$ of this model.
  This corresponds to an increse in pairing strengths by $\sim$10\%.
 We decided to scale proton pairing strengths by the same factor 1.1
 to preserve the original $G_p/G_n$ ratio of the MM model.
 Such stronger pairing produces smaller nuclear masses (i.e., increases
 binding) by ~3.5 - 4.5 MeV for studied nuclei.
 One could try to compensate for this change by subtracting an increased
 average pairing energy term, but we did not attempt any modification of our
 mass model. The sole aim of using PNP was to obtain a more realistic estimate
 for energies of 3q.p. excitations.

%  The calculation without blocking is much simpler and we were
% able to perform a seven-dimensional minimization over axially-symmetric
% deformations
% $\beta_2, \beta_3, \beta_4, \beta_5, \beta_6, \beta_7$ and $\beta_8$.
% Therefore, these results should be reliable also for light actinides.
% As we preferred to avoid a new fit of the macroscopic model parameters,
% also for this model we introduced three additive constants (energy shifts)
% for even-odd,  odd-even and odd-odd nuclei which
%  minimize the rms deviation in each of the groups of nuclei.

Equilibrium deformations were found by both the blocked BCS and quasiparticle methods for more than 2500 one-proton and two-neutron
($1\pi2\nu$) and for more than 500 3-proton ($3\pi$) 3-q.p. configurations built from s.p.
states not too distant from the Fermi surfaces in studied nuclei. The PNP calculations were confined to some selected low-lying 3q.p. states.
Configurations at the lowest excitation energies $E_{3q.p.}^*(K)$ are
considered as likely candidates for $K$-isomers. We did not consider
 shifts due to the spin interaction (counterpart of Gallagher shifts for
2q.p.), which for 3q.p. configurations are not well studied
\cite{Pyatov1964,Jain1992}.

Since a 3-q.p. configuration in an odd-$Z$ deformed nucleus may be thought of
as a 2-neutron (or a 2-proton) excitation built on its 1-proton component which is a band-head of some rotational band, the probability
of deexcitation to this band is one of the factors determining the isomerism. For a $1\pi2\nu$ configuration
with $K=\Omega_{\pi}+\Omega_{\nu 1}+\Omega_{\nu 2}$, its excitation energy over the 1-proton band-head with
$K_{\pi}=\Omega_{\pi}$, $E_{3q.p.}^*(K)-E_{1q.p.}^*(\Omega_{\pi})$, should be compared to the collective rotational energy:
$E^{rot}(I=K,\Omega_{\pi})=\frac{1}{2}[K(K+1)-\Omega_{\pi}^2]/{\cal J}=
      \frac{1}{2}[K_{\nu}(K_{\nu}+2\Omega_{\pi}+1)+\Omega_{\pi}]/{\cal J} $,
with $K_{\nu}=\Omega_{\nu 1}+ \Omega_{\nu 2}$, and ${\cal J}$ - average moment of inertia of the 1-proton q.p. rotational band.
For $3\pi$ configurations with $K=\sum_1^3\Omega_{\pi i}$ similar comparisons could be made.
The value $(E_{3q.p.}^*(K)-E_{1q.p.}^*(\Omega_{\pi}))-E^{rot}(I=K,\Omega_{\pi})$ gives some indication
of a likelihood of the 3-q.p. configuration being isomeric: the smaller it is the less probable is the high-$K$ state
deexcitation to the one-proton q.p. rotational band.
%  criterion but now there are three possibilieties. Depending on which
%  one-proton state $\Omega_{\pi i}$ is taken as the rotational band-head, one
%  has to replace $K_{\nu}$ with $\sum_{k\ne i} \Omega_{\pi k}$ in the
%  abovementioned formula. Which of the three conditions is the most stringent
%  will depend on both, $\Omega_{\pi i}$ and excitation energies of 1 (quasi-)
% proton configurations.

Unfortunately, the above energy difference is not a precise indicator of the $K$-isomerism. A customary
indicator - the reduced hindrance $f_\nu$, is based on the knowledge of the $EM$ transition depopulating the high-$K$ configuration
and its final state: $f_\nu=F^{1/\nu}$, where $F=\tau^{\gamma}/\tau^W$ is the ratio of the partial $EM$ half-life to its s.p. Weisskopf
estimate, $\nu=\Delta K-\lambda$, with $\Delta K$ the difference between the initial and final state $K$-values, and $\lambda$ - transition
multipolarity. As can be seen from the experimental data \cite{kon2,Dracoulis2016}, the high-$K$ isomers in deformed nuclei can
occur at substantial excitation energies above the yrast line. For example, 2-q.p. $K^{\pi}=8^-$ isomers in even-even nuclei from various deformed
regions, shown in Fig. 12 in \cite{kon2}, which occur 0.5 - 1.0 MeV above the yrast line still have substantially hindered decays
with $f_{\nu}$ values above 30 (typically, $f_\nu=30$-200 for isomers). Thus, by analogy, 3-q.p. configurations characterized
by the energy differences $(E_{3q.p.}^*(K)-E_{1q.p.}^*(\Omega_{\pi}))-E^{rot}(I=K,\Omega_{\pi})
\lesssim 1$ MeV (which are the counterparts of the excitation energies above the yrast line displayed in \cite{kon2}) can be
considered as candidates for isomers, those with smaller differences being preferable.
Notice that
%if moments of inertia ${\cal J}$ of all rotational bands built on
% 1 q.p.  proton excitations in a given nucleus would be equal, one would have:
%  $E^{rot}(I+\Omega_{\pi 2},\Omega_{\pi 2})-
% E^{rot}(I+\Omega_{\pi 1},\Omega_{\pi 1})=(\Omega_{\pi 2}-\Omega_{\pi 1})
% (2I+1)/(2{\cal J})$, so that
the rotational energy of levels of a 1-q.p. band grows with increasing $\Omega_{\pi}$. For example, with $2{\cal J}=170\ \hbar^2$/MeV,
the rotational energy at angular momentum $\Omega_{\pi}+I_R$ with $I_R=6$ and
11  amounts to 285 and 670 keV, respectively, for $\Omega_{\pi}=1/2$, and to 850 and 1530 keV for $\Omega_{\pi}=11/2$. Hence, with ${\cal J}$ being
similar for various 1-proton rotational bands, a 3-q.p. configuration containing the lowest proton (i.e. g.s.) orbital with a larger
$\Omega_{\pi}$ has a greater chance to be isomeric.

To estimate rotational energies in studied isotopes we used the calculated cranking moments of inertia of even-even nuclei
from \cite{momJ}. For odd-$Z$ nuclei, we took the average from calculated moments of inertia in neighboring even-even nuclei and
increased it by a factor accounting for two effects: overall larger moments of inertia in odd-$A$ vs even-even nuclei as seen in
actinides, and the observed increase in ${\cal J}$ with rotational frequency (or collective angular momentum) above the cranking value
for spin-zero which was given in \cite{momJ}. For our estimates, we arbitrarily used the factor 1.4. Clearly, an increase in the moment of
inertia of the g.s. rotational band makes 3-q.p. high-$K$ configurations more
 excited with respect to it.

\section{RESULTS AND DISCUSSION}

%\subsection{Excitation energies of 3-q.p. states }

Calculated g.s. deformations in considered nuclei change according to the following pattern. The deformations $\beta_{2 0}$ are mostly between
0.20 and 0.25, with the largest values for $N=148-158$, slightly decreasing for $N\geq 160$ and with increasing $Z$.
The deformations $\beta_{4 0}$ decrease with $N$ by $\sim 0.10$ from positive to negative values, starting from $\beta_{4 0}\approx 0.06$
for $Z=101, N=142$, and from $\beta_{4 0}\approx 0.02$ for  $Z=111, N=148$. The deformations $\beta_{6 0}$ are mostly negative with the
largest magnitude for $N=150,152$: $\beta_{6 0} \approx -0.06$ for $Z=101, N=150$, and $\beta_{6 0}\approx -0.03$ for $Z=111, N=150$.
Finally, the deformations $\beta_{8 0}$ are generally small, with the largest values $\beta_{8 0}\approx 0.03$ for $N=158,160$.
Equilibrium deformations for the majority of 3-q.p. configurations are close to those of the ground states, which in the case of $\beta_{20}$ means
that it falls within the range $\pm 0.02$ around the ground-state value.

The Woods-Saxon single-particle spectra in Lr isotopes at the calculated g.s. deformations are shown in Fig. \ref{p103} and \ref{n103}.
The proton s.p. states with large $\Omega$ that can form high-$K$ configurations are (from bottom to top):
% states for both nuclei: $9/2^{-}[734]$; $\nu 7/2^{-}[624]$; $\nu 5/2^{+}[622]$.
%%%%%%%%%%%%%%%%%%%%%%%%%%%%%%%%%%%%%%
\begin{figure}[!tbp]
\centerline{\includegraphics[scale=0.8]{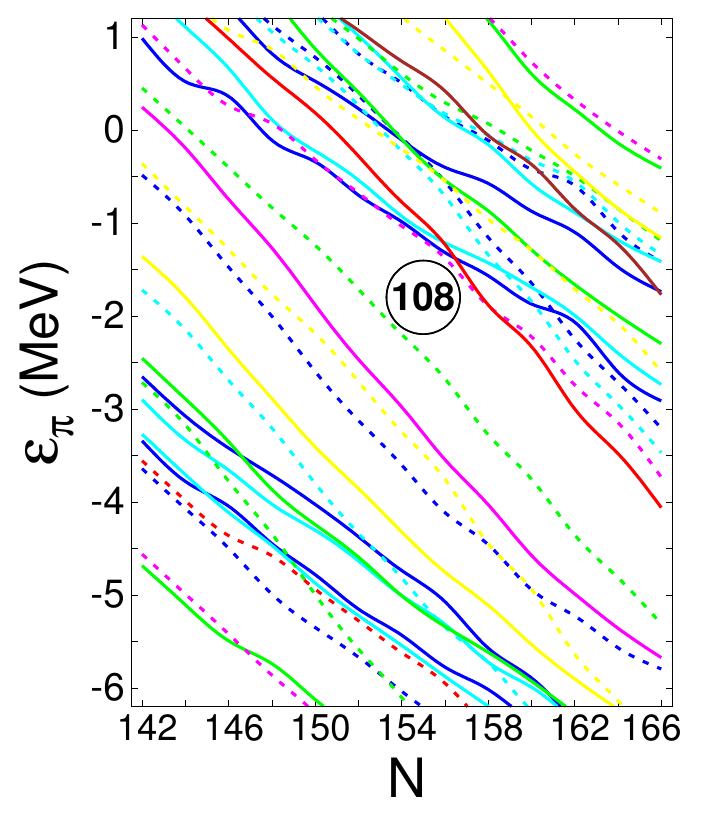}}
\caption{{\protect Single-particle proton levels for Lr isotopes at their equilibrium deformations; negative
 parity states - dashed, positive parity - solid lines; $\Omega=1/2$ - blue,
 $3/2$ - cyan, $5/2$ - green, $7/2$ - yellow, $9/2$ - magenta, $11/2$ -
 red, $13/2$ - brown.
 %Equilibrium deformations as given in Tables \cite{}.
 }}
\label{p103}
\end{figure}
%%%%%%%%%%%%%%%%%%%%%%%%%%%%%%%%%%%%%%
%%%%%%%%%%%%%%%%%%%%%%%%%%%%%%%%%%%%%%
\begin{figure}[!tbp]
\centerline{\includegraphics[scale=0.8]{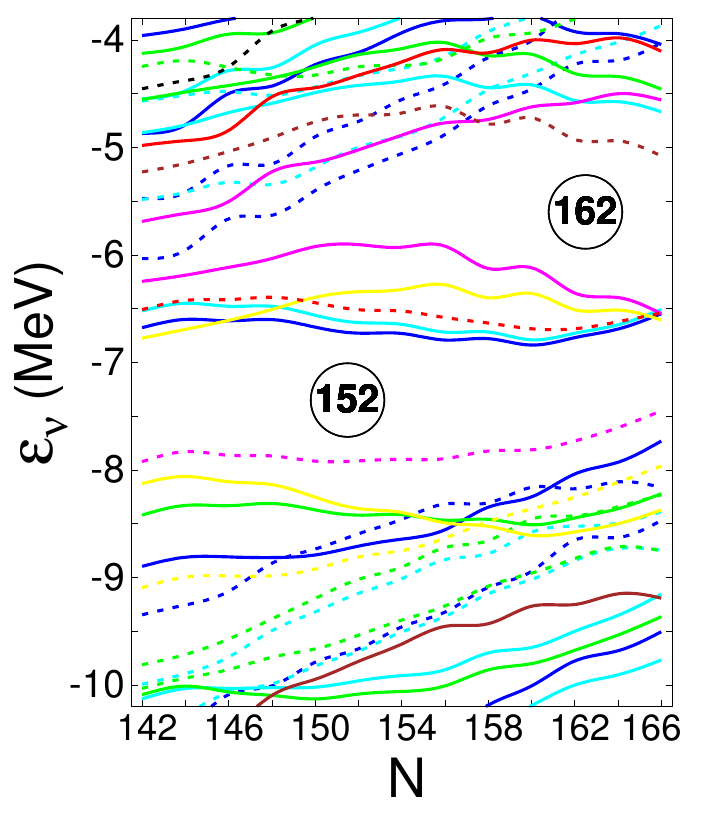}}
\caption{{\protect Single-particle neutron levels for Lr isotopes at their equilibrium deformations.; negative
 parity states - dashed, positive parity - solid lines; $\Omega=1/2$ - blue,
 $3/2$ - cyan, $5/2$ - green, $7/2$ - yellow, $9/2$ - magenta, $11/2$ -
 red, $13/2$ - brown, $15/2$ - black.
% Equilibrium deformations as given in Tables \cite{}.
 }}
\label{n103}
\end{figure}
%%%%%%%%%%%%%%%%%%%%%%%%%%%%%%%%%%%%%%
%\begin{figure}[h]
%\centerline{\includegraphics[scale=0.6]{n109.pdf}}
%\caption{{\protect Single - particle neutron levels for Mt isotopes; negative
% parity states - dashed, positive partity - solid lines; $\Omega=1/2$ - black,
% $3/2$ - blue, $5/2$ - green, $7/2$ - green-yellow, $9/2$ - yellow, $11/2$ -
% red, $13/2$ - purple, $15/2$ - cyan. Equilibrium deformations as given in
% Tables \cite{}.}}
%\label{n109}
%\end{figure}
%\begin{figure}[h]
%\centerline{\includegraphics[scale=0.6]{p109.pdf}}
%\caption{{\protect Single - particle proton levels for Mt isotopes; negative
% parity states - dashed, positive partity - solid lines; $\Omega=1/2$ - black,
% $3/2$ - blue, $5/2$ - green, $7/2$ - green-yellow, $9/2$ - yellow, $11/2$ -
% red, $13/2$ - purple. Equilibrium deformations as given in
% Tables \cite{}.}}
%\label{p109}
%\end{figure}
$\pi 7/2^{+}_{\sf 3} [633]$, $\pi 7/2^{-}_{\sf 3} [514]$, $\pi 9/2^{+}_{\sf 2} [624]$, $\pi 9/2^{-}_{\sf 2} [505]$,
$\pi 11/2^{+}_{\sf 1} [615]$ and $\pi 7/2^{-}_{\sf 4} [503]$. S.p. states are labeled by $\Omega^{\pi}_n$, with $n$ - the number of the state
(counted from the lowest one) within the $\Omega^{\pi}$ block. The provided Nilsson labels serve to make a connection to the traditional scheme;
they have no or little sense for lower $\Omega$ values due to their mixing in realistic potentials. Nevertheless, due to the widespread use of Nilsson's notation in the works of other authors,
we also decided to use it, mostly in figures, alternately with the  $\Omega^{\pi}_n$-one.
% The the same plus addittional $\pi 3/2^{-}[512]$ for $N>158$ in Rg.

 Proton states at the Fermi level in considered odd-$Z$ nuclei determine their
 g.s. spins and parities. In the quasiparticle
scheme, these are in Md: $1/2^-_{\sf 10}$ ([521] in the Nilson scheme) with the exception
of $7/2^-_{\sf 3}$ for $N=160$ and $7/2^+_{\sf 3}$ for $N=166$; in Lr: $7/2^-_{\sf 3}$ except for $9/2^+_{\sf 2}$ for $N=160,162,166$
and $1/2^-_{\sf 10}$ for $N=164$ (for $N=$160-166 states $9/2^+_{\sf 2}$ and $1/2^-_{\sf 10}$ are practically
degenerate); in Db: $9/2^+_{\sf 2}$, except for $5/2^-_{\sf 5}$ for $N=$164,166; in Bh: $5/2^-_{\sf 5}$; in Mt:
$9/2^-_{\sf 2}$ for $N=146-152$ and $11/2^+_{\sf 1}$ for $N=$154-166; in Rg:
$9/2^-_{\sf 2}$ for $N=$152,154, $3/2^-_{\sf 8}$ for $N=$162-166 and
$11/2^+_{\sf 1}$ for other isotopes, with two high-$\Omega$ states being
nearly degenerate for $N=$150, 154, 156. The g.s. spins and parities from the PNP calculation differ from
the above only in a few cases.

From a comparison to the experimentally established spins of low-lying states in Md and Lr isotopes around $N=150$
it follows that the order of proton $\pi 1/2^{-}_{\sf 10}$ and $\pi 7/2^{-}_{\sf 3}$ states is inverted in the Woods-Saxon potential.

In view of the g.s. deformation changing with $Z$ we have to consider a greater number of two-neutron combinations than
necessary for only one isotopic chain. Relevant large - $\Omega$ neutron states are (going from bottom to top):
$\nu 7/2^{-}_{\sf 5} [743]$, $\nu 7/2^{+}_{\sf 4} [624]$,
$\nu 9/2^{-}_{\sf 3} [734]$, $\nu 7/2^{+}_{\sf 5} [613]$,
$\nu 11/2^{-}_{\sf 2} [725]$,
$\nu 9/2^{+}_{\sf 3} [615]$, $\nu 9/2^{+}_{\sf 4} [604]$ and
$\nu 13/2^{-}_{\sf 1} [716]$. These states enter the neutron 2-q.p. component of the lowest 3-q.p. configurations in the considered region of nuclei.
%listed in Table.

%\begin{figure}[h]
%\centerline{\includegraphics[scale=0.8]{spektrum_249Md.pdf}}
%\caption{{\protect Single particle spectrum of $^{249}$ Md for protons states (left) and
%neutrons states (right).}}
%\label{spektrum_249Md}
%\end{figure}

%In $^{251}$Md state $9/2^{-}[734]$ basically coincides exactly
%with ($\lambda_{n}$) while $\nu 7/2^{-}[624]$ and $\nu 5/2^{+}[622]$ in given order are located under it.

%\begin{figure}[h]
%\centerline{\includegraphics[scale=0.8]{spektrum_251Md.pdf}}
%\caption{{\protect Single particle spectrum of $^{251}$ Md for protons states (left) and
%neutrons states (right).}}
%\label{spektrum_251Md}
%\end{figure}

\subsection{Excitation energies of $1\pi2\nu$ 3-q.p. \mbox{high-$K$} -
 candidates for isomeric states}

\begin{figure*}
\includegraphics[width=1.1\linewidth]{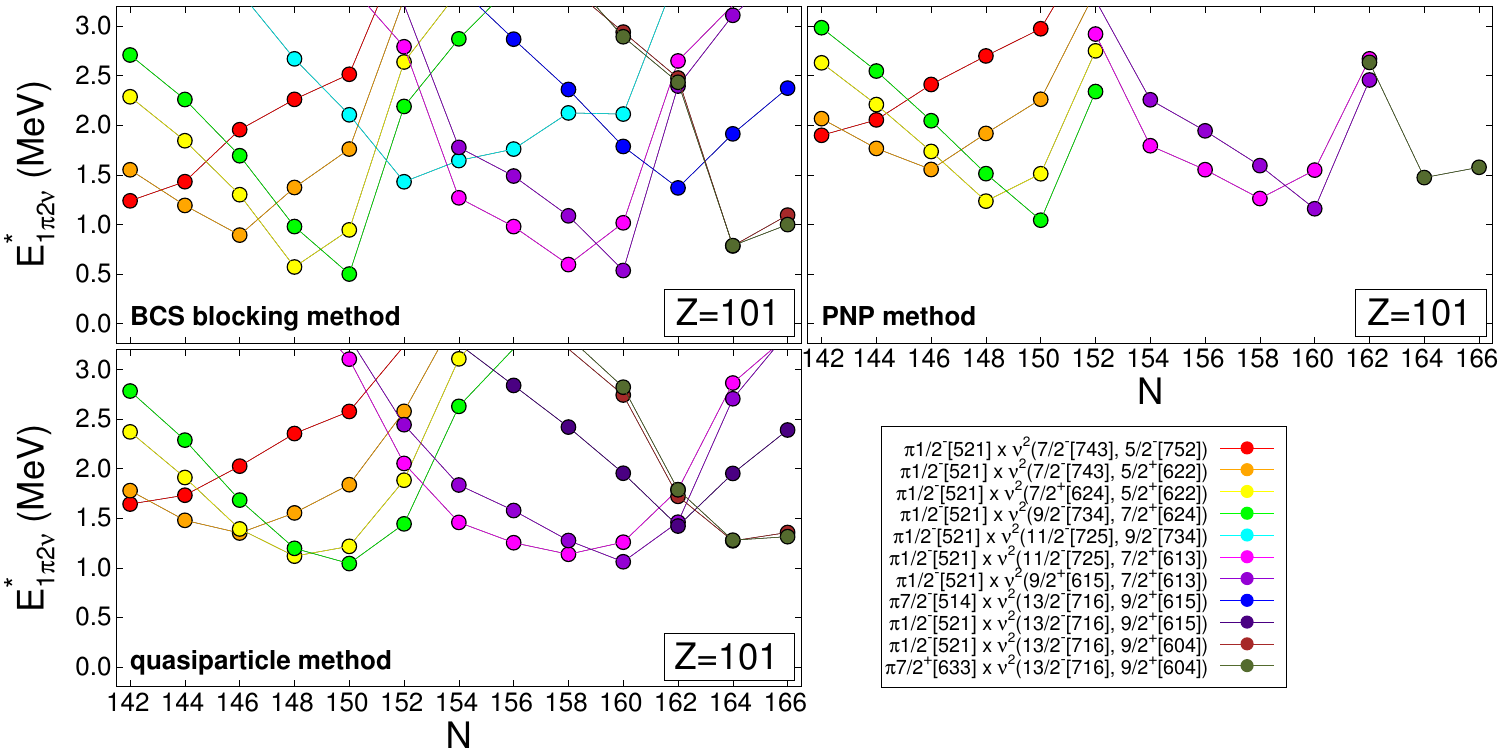}
\caption{Calculated excitation energy vs $N$ for $1\pi2\nu$ large-$K$ configurations which for some $N$ become
 the lowest in odd-even isotopes of Md; each panel corresponds to the indicated method of calculations.}
\label{1p2n_101}
\end{figure*}

\begin{figure*}
\includegraphics[width=1.1\linewidth]{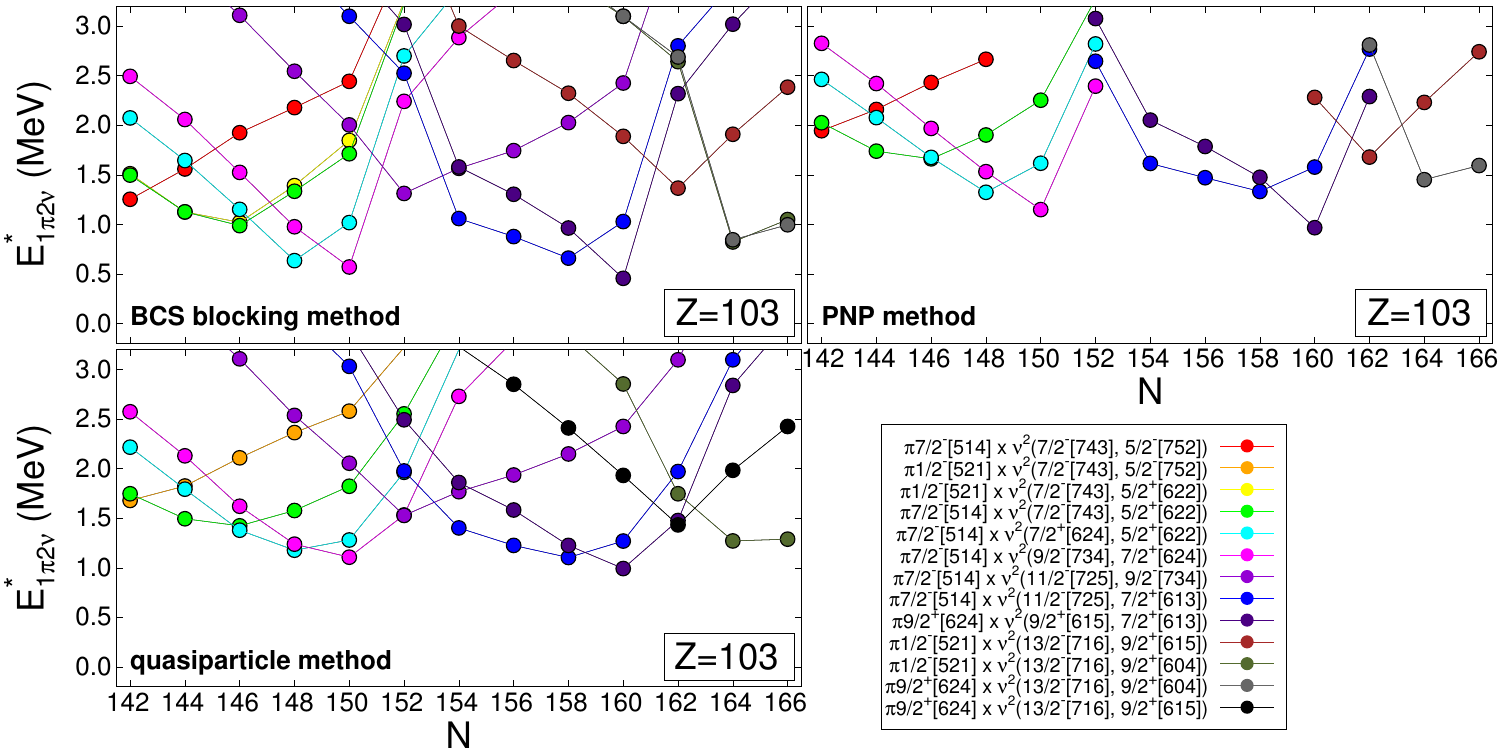}
\caption{ As in Fig. \ref{1p2n_101} but for Lr.}
\label{1p2n_103}
\end{figure*}

\begin{figure*}
\includegraphics[width=1.1\linewidth]{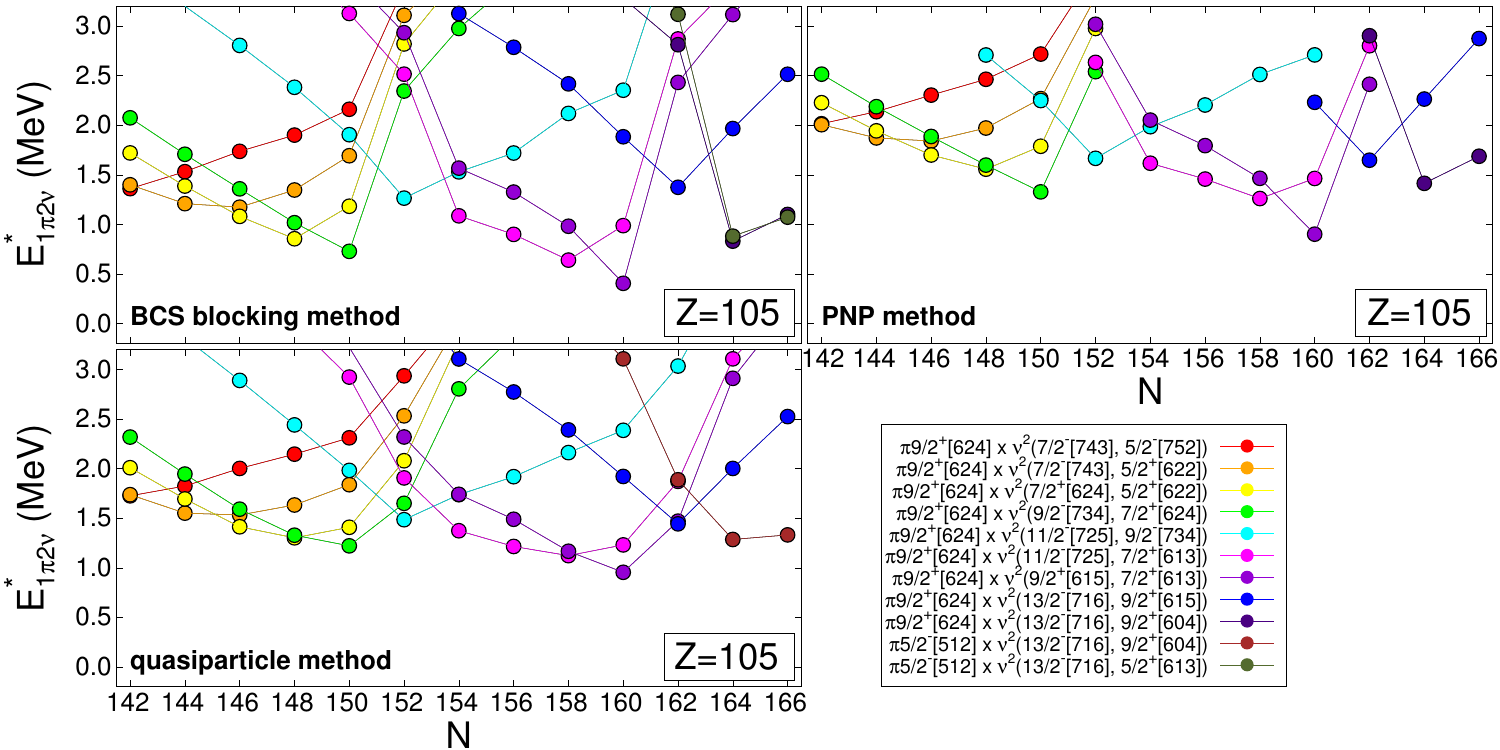}
\caption{As in Fig. \ref{1p2n_101} but for Db.}
\label{1p2n_105}
\end{figure*}

\begin{figure*}
\includegraphics[width=1.1\linewidth]{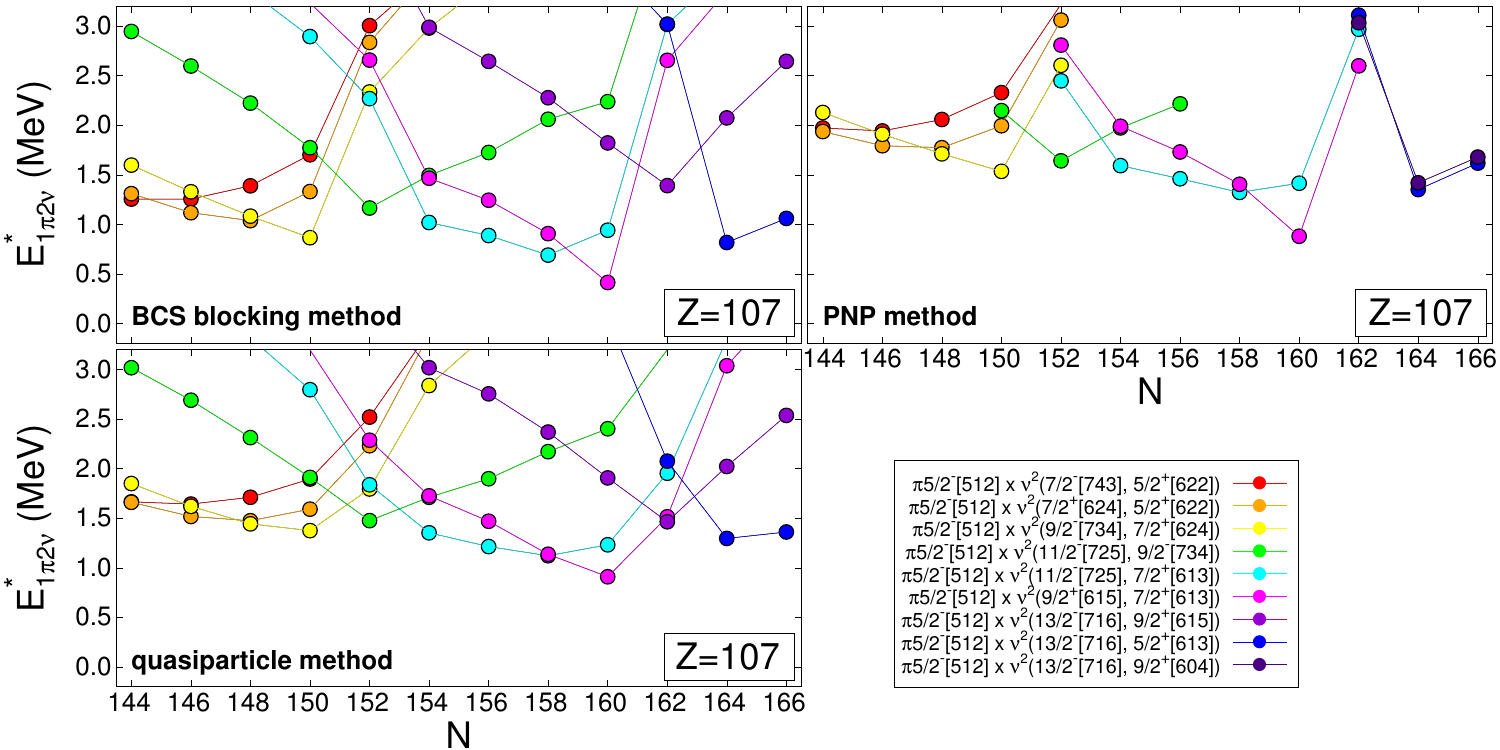}
\caption{As in Fig. \ref{1p2n_101} but for Bh.}
\label{1p2n_107}
\end{figure*}

\begin{figure*}
\includegraphics[width=1.1\linewidth]{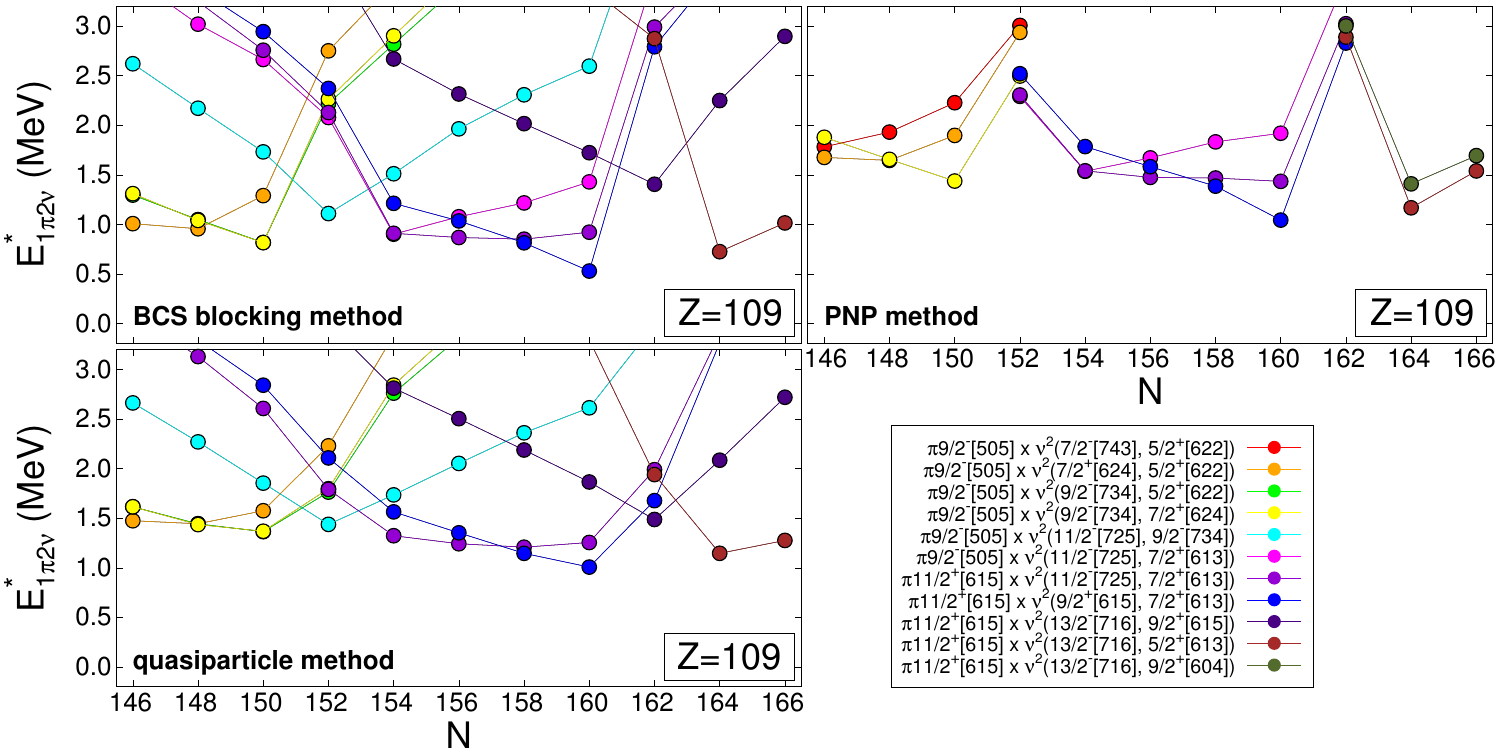}
\caption{As in Fig. \ref{1p2n_101} but for Mt.}
\label{1p2n_109}
\end{figure*}

\begin{figure*}
\includegraphics[width=1.1\linewidth]{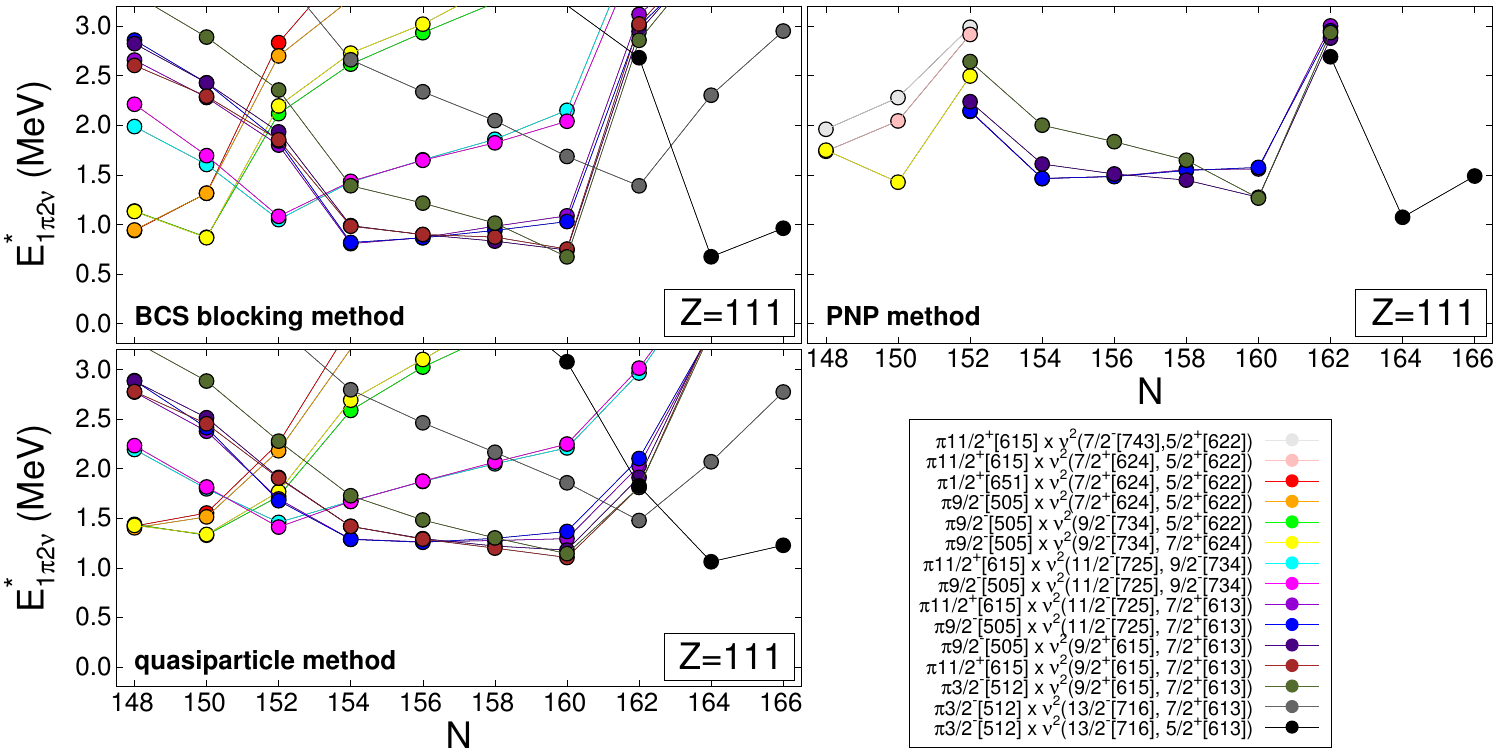}
\caption{As in Fig. \ref{1p2n_101} but for Rg.}
\label{1p2n_111}
\end{figure*}

From calculated excitation energies of $1\pi2\nu$ large-$K$ states for six
 odd-even isotopic chains we select those with the lowest energies at some $N$.
In Fig. \ref{1p2n_101} - \ref{1p2n_111} are shown such candidates and
 corresponding excitation energies $E_{1\pi2\nu}^{*}$ obtained within
the standard BCS method with blocking (top left panels), quasiparticle method
 (bottom left panels), and from the PNP calculation (for selected
 configurations - top right panels).
Tabulated results for five lowest-lying configurations in each isotope
 obtained in the blocked BCS and quasiparticle method are provided in the
 supplement material.

%%%%%%%%%%%%%%%%%%%%%%%%%%%%%%%%%%%%%%%%%%%%%%%%%%%%%%%%%%%%%%%%%%%%%%%%%%%%%%%%%%%%%%%%%%%%%%%%%%%

%%%%%%%%%%%%%%%%%%%%%%%%%%%%%%%%%%%%%%%%%%%%%%%%%%%%%%%%%%%%%%%%%%%%%%%%%

We rely mostly on pairing calculations within the quasiparticle and PNP schemes, as those with the BCS
blocking give too small excitation energies of 3 q.p. configurations. This is expected as BCS solutions with
pairing strength adjusted to the g.s. produce too weak correlations or even unpaired solutions when
two or three levels are blocked.
Results of all three pairing schemes point to the same configurations which we review below. Generally, the isotopic variation of excitation
energies is milder within the quasiparticle method than within the PNP and BCS blocking methods.

With a changing neutron number various low-laying 2-neutron configurations occur in specific isotopes.
One can divide those leading to particularly low-lying high-$K$ states into three groups: type A) in $N=148-150$ isotopes, type B) in $N=152-160$ isotopes,
and type C) in $N=164-166$ isotopes. Those energetically favored among others in specific isotopes are: of type A) -
 $\nu^2 6^+ \{\nu 7/2^-_{\sf 5}\otimes \nu 5/2^-_{\sf 7}\} $,
 $\nu^2 6^- \{\nu 7/2^-_{\sf 5}\otimes \nu 5/2^+_{\sf 7}\} $,
 $\nu^2 6^+ \{\nu 7/2^+_{\sf 4}\otimes \nu 5/2^+_{\sf 7}\} $, and
 $\nu^2 8^- \{\nu 7/2^+_{\sf 4}\otimes \nu 9/2^-_{\sf 3}\} $; of type B) -
 $\nu^2 9^- \{\nu 7/2^+_{\sf 5}\otimes \nu 11/2^-_{\sf 2}\} $,
 $\nu^2 8^+ \{\nu 7/2^+_{\sf 5}\otimes \nu  9/2^+_{\sf 3}\} $, and
 $\nu^2 10^+ \{\nu 9/2^-_{\sf 3}\otimes \nu 11/2^-_{\sf 2}\} $; of type C) -
 $\nu^2 11^+ \{\nu 9/2^-_{\sf 3}\otimes \nu 13/2^-_{\sf 1}\} $,
 $\nu^2 11^- \{ \nu 9/2^+_{\sf 4}\otimes \nu 13/2^-_{\sf 1}\} $, and
 $\nu^2 9^- \{\nu 5/2^+_{\sf 8}\otimes \nu 13/2^-_{\sf 1}\} $.
At the same time, the Woods-Saxon single-neutron level scheme leads to
the lowest $1\pi2\nu$ excitations
in $N=152$ and $N=162$ isotones lying higher than in the neighboring ones. This effect is weakened by the very small or vanishing BCS neutron gap in the quasiparticle method, but is very prominent with the PNP and the blocked BCS. Below we discuss results for separate isotopic chains, mostly from the quasiparticle scheme.
The results of the PNP scheme will be commented on at the end of this subsection.

%%%%%%%%%%%%%%%%%%%%%%%%%%%%%%%%%%%%%%%%%%%%%%%%%%%%%%%%%%%%%%%%%%%%%%%%%%%%%%%%%%%%%%%%%%%%%%%%%%%

{\it Md\ \ } The lowest-lying type-A) configurations are $\pi \nu^2 13/2^{-}\{ \pi 1/2^{-}_{\sf 10} \otimes \nu 7/2^{+}_{\sf 4}
\otimes \nu 5/2^{+}_{\sf 7}\}$ (yellow dots in Fig. \ref{1p2n_101}) in $^{249}$Md
and $\pi \nu^2 17/2^{+}\{ \pi 1/2^{-}_{\sf 10} \otimes \nu 9/2^{-}_{\sf 3} \otimes
 \nu 7/2^{+}_{\sf 4}\}$ (green dots) in $^{251}$Md. The latter, at 1.04 MeV above
 the g.s., is the lowest-lying type A) state of all.
 The $\pi\nu^2 15/2^{+}\{ \pi 1/2^{-}_{\sf 10} \otimes \nu 9/2^{-}_{\sf 3}
 \otimes \nu 5/2^{+}_{\sf 7}\}$ configuration is the second lowest,
 at similar excitation energy in both isotopes. Although configurations
 with $\pi 7/2^{-}_{\sf 3}$ orbital replacing $\pi 1/2^-_{\sf 10}$ lie by
 0.25 MeV higher they should be considered in view of the seemingly
 opposite order of both states in the experiment as compared to the present WS
 spectrum.
%\begin{figure}[h]
%\centerline{\includegraphics[scale=0.7]{blokowanie.pdf}}
%\caption{{\protect Excitation energy for mendelevium in the neutron range from N=142 up to N=156  within blocking method.}}
%\label{blokowanie}
%\end{figure}
 In $N=154, 156, 158$ isotopes the lowest-lying type-B) configuration is
 $\pi \nu^2 19/2^+ \{\pi 1/2^-_{\sf 10}\otimes \nu 7/2^+_{\sf 5} \otimes \nu 11/2^-_{\sf 2}\} $ (magenta dots);
 $\pi \nu^2 17/2^- \{\pi 1/2^-_{\sf 10}\otimes \nu 7/2^+_{\sf 5} \otimes \nu 9/2^+_{\sf 3}\} $ (dark violets dots) is the
 lowest one in $^{261}$Md; the configurations with $\pi 7/2^-_{\sf 3}$ and
 $\pi 9/2^+_{\sf 2}$ replacing $\pi 1/2^-_{\sf 10}$ are lying 100-200 keV
 higher, but the first one may be relevant if the proton level order established in $N\approx 150$ isotopes persists in the heavier ones.
 The type-C) configurations including
 the 2-neutron pair $\{\nu 9/2^+_{\sf 4} \otimes \nu 13/2^-_{\sf 1}\}$ and one of:
 $\pi 1/2^-_{\sf 10}$, $\pi 7/2^+_{\sf 3}$ (black dots in Fig.
 \ref{1p2n_101}) or $\pi 7/2^-_{\sf 3}$, are the favoured ones in $^{265}$Md
 (the two first configurations have nearly the same energy),
 while the ones with the same neutron contents and either $\pi 7/2^+_{\sf 3}$
 or $\pi 9/2^+_{\sf 2}$ are the lowest ones in $^{267}$Md.

%%%%%%%%%%%%%%%%%%%%%%%%%%%%%%%%%%%%%%%%%%%%%%%%%%%%%%%%%%%%%%%%%%%%%%%%%%%%%%%%%%%%%%%%%%%%%%%%%%%

 Estimated excitation energies of the lowest $1\pi2\nu$ configurations above the
 rotational g.s. structure, discussed in Sect. II, are: in $N=150$
 ($K_\nu=8$): $\approx 0.6$ MeV for
 $\Omega_{\pi}=1/2$ and $\approx 0.35$ MeV for $\Omega_{\pi}=7/2$; in $N=154$
 ($K_{\nu}=9$): $\approx 0.95$ MeV for $\Omega_{\pi}=1/2$ and $\approx 0.65$
 for $\Omega_{\pi}=7/2$; in $N=158$ ($K_{\nu}=9$): $\approx 0.55$ for
 $\Omega_{\pi}=1/2$ and $\approx 0.25$ for $\Omega_{\pi}=7/2$; in $N=160$
 ($K_{\nu}=8$): $\approx 0.6$ MeV for $\Omega_{\pi}=1/2$ and $\approx 0.3$ MeV
 for $\Omega_{\pi}=7/2$. We have not calculated the moment of inertia for
  $N=164$ Md, but assuming that for the Lr isotone one obtains ($K_{\nu}=11$):
  $\approx 0.45$ for $\Omega_{\pi}=1/2$ and $\approx -0.25$ MeV (the yrast
 trap) for $\Omega_{\pi}=7/2$ (the $K$-mixing of nearly degenerate
 proton levels would probably remove the trap effect).

 {\it Lr\ \ } The lowest-lying $1\pi2\nu$ type-A) configurations are:
 $\pi \nu^2 19/2^- \{\pi 7/2^-_{\sf 3} \otimes \nu 7/2^+_{\sf 4} \otimes \nu 5/2^+_{\sf 7}\} $ in $^{251}$Lr
  and
  $\pi \nu^2 23/2^+ (25/2^-) \{\pi 7/2^-_{\sf 3} (\pi 9/2^+_{\sf 2})
 \otimes\nu 9/2^-_{\sf 3}\otimes\nu 7/2^+_{\sf 4} \} $ or $\pi \nu^2 21/2^+
\{\pi 7/2^-_{\sf 3}\otimes\nu 9/2^-_{\sf 3}\otimes \nu 5/2^+_{\sf 7}\} $ in
 $^{253}$Lr (Fig.\ref{1p2n_103}).
 Configurations with
 $\pi 1/2^-_{\sf 10}$ replacing $\pi 7/2^-_{\sf 3}$ are lying slightly higher.
 The lowest type B) states are:
 $\pi \nu^2 25/2^+ \{\pi 7/2^-_{\sf 3}\otimes \nu 11/2^-_{\sf 2}\otimes \nu 7/2^+_{\sf 5}\} $ in $^{261}$Lr and
 $\pi \nu^2 25/2^+ \{\pi 9/2^+_{\sf 2}\otimes \nu 9/2^+_{\sf 3}\otimes \nu 7/2^+_{\sf 5}\} $ in $^{263}$Lr;
 the configurations with interchanged proton states are predicted as the second
 lowest in both nuclides.
 Although not very low-lying, the type C), large-$K$ configurations:
 $\pi \nu^2 23/2^+ \{\pi 1/2^-_{\sf 10} \otimes \nu 9/2^+_{\sf 4} \otimes
 \nu 13/2^-_{\sf 1}\} $ or $\pi \nu^2 31/2^- \{\pi 9/2^+_{\sf 2} \otimes \nu 9/2^+_{\sf 4} \otimes
 \nu 13/2^-_{\sf 1}\} $ may be good candidates in $^{267,269}$Lr.

%%%%%%%%%%%%%%%%%%%%%%%%%%%%%%%%%%%%%%%%%%%%%%%%%%%%%%%%%%%%%%%%%%%%%%%%%%%%%%%%%%%%%%%%%%%%%%%%%%%

%%%%%%%%%%%%%%%%%%%%%%%%%%%%%%%%%%%%%%%%%%%%%%%%%%%%%%%%%%%%%%%%%%%%%%%%%%%%%%%%%%%%%%%%%%%%%%%%%%%

  {\it Db\ \ } The most promising candidates occur in the heavier isotopes
 (type B),
 $\pi\nu^2 27/2^-\{\pi 9/2^+_{\sf 2}\otimes
 \nu 11/2^-_{\sf 2}\otimes \nu 7/2^+_{\sf 5}\}$, $\pi\nu^2 25/2^+\{7/2^-_{\sf 3}\otimes
 \nu 11/2^-_{\sf 2}\otimes \nu 7/2^+_{\sf 5}\}$ in $^{263}$Db and $\pi\nu^2 25/2^+ (21/2^-, 23/2^-) \{\pi 9/2^+_{\sf 2} (5/2^-_{\sf 5},7/2^-_{\sf 3}) \otimes
 \nu 9/2^+_{\sf 3}\otimes \nu 7/2^+_{\sf 5}\}$
 %$\pi\nu^2 23/2^- \{\pi 7/2^-_{\sf 3} \otimes
 %\nu 9/2^+_{\sf 3}\otimes \nu 7/2^+_{\sf 5}\}$
  in $^{265}$Db  (Fig. \ref{1p2n_105}).
 The type A) low-lying configurations are:
 $\pi\nu^2 25/2^-\{\pi 9/2^+_{\sf 2}\otimes \nu 9/2^-_{\sf 3}\otimes
\nu 7/2^+_{\sf 4}\}$ and
 $\pi\nu^2 23/2^-\{\pi 9/2^+_{\sf 2}\otimes \nu 9/2^-_{\sf 3}\otimes
 \nu 5/2^+_{\sf 7}\}$ in $^{255}$Db, and
 $\pi\nu^2 21/2^+\{\pi 9/2^+_{\sf 2}\otimes \nu 5/2^+_{\sf 7}\otimes
 \nu 7/2^+_{\sf 4}\}$ in $^{253}$Db.
 Type-C) candidates with sizable $K$ are:
 $\pi\nu^2 31/2^+\{\pi 9/2^+_{\sf 2} \otimes
 \nu 13/2^-_{\sf 1}\otimes \nu 9/2^+_{\sf 4}\}$, $\pi\nu^2 27/2^-\{\pi 5/2^-_{\sf 5} \otimes \nu 13/2^-_{\sf 1}\otimes \nu 9/2^+_{\sf 4}\}$ in $^{269}$Db and
 the same neutron pair coupled to $\pi 5/2^-_{\sf 5}$ or $\pi 1/2^-_{\sf 10}$
 proton states in $^{271}$Db.
 Their estimated excitation above the rotational
 sequence based on the 1-proton component is close to zero for $\Omega_{\pi}=5/2$
 and even less than zero for $\Omega_{\pi}=9/2$. Again, since these states
 have very similar energies one can expect some $K$-mixing.

%
%%%%%%%%%%%%%%%%%%%%%%%%%%%%%%%%%%%%%%%%%%%%%%%%%%%%%%%%%%%%%%%%%%%%%%%%%%%%%%%%%%%%%%%%%%%%%%%%%%%
 {\it Bh\ \ } The lowest-lying candidate for isomer of all $1\pi2\nu$ 3-q.p.
 states is predicted in $^{267}$Bh ($N=160$), in which the configuration
 $\pi\nu^2 21/2^-\{\pi 5/2^-_{\sf 5}\otimes \nu 9/2^+_{\sf 3} \otimes
 \nu 7/2^+_{\sf 5}\}$ lies at 910 keV (see Fig. \ref{1p2n_107}), at an
 estimated 0.3 MeV above the rotational g.s. band;
  the next one, $\pi\nu^2 25/2^+\{\pi 5/2^-_{\sf 5}\otimes \nu 9/2^+_{\sf 3}
 \otimes \nu 11/2^-_{2}\}$, with $K$ bigger by two units, lies already
  300 keV higher.
 The configuration: $\pi\nu^2 23/2^+\{\pi 5/2^-_{\sf 5}\otimes
 \nu 11/2^-_{\sf 2} \otimes \nu 7/2^+_{\sf 5}\}$
 is the lowest in $^{263,265}$Bh (this in the $N=158$ isotone lying 100
 keV lower). Type-A) configurations:
 $\pi\nu^2 21/2^+ \{\pi 5/2^-_{\sf 5}\otimes \nu 9/2^-_{\sf 3} \otimes \nu 7/2^+_{\sf 4}\}$, $\pi\nu^2 19/2^+\{\pi 5/2^-_{\sf 5}\otimes \nu 9/2^-_{\sf 3} \otimes
 \nu 5/2^+_{\sf 7}\}$ in $^{257}$Bh have excitation energy
 larger by more than 300 keV than the corresponding ones in Md $N=150$ isotone.
  Type C) configurations:
 $\pi\nu^2 23/2^+ \{\pi 5/2^-_{\sf 5}\otimes \nu 13/2^-_{\sf 1} \otimes
 \nu 5/2^+_{\sf 8}\}$, $\pi\nu^2 27/2^+ \{\pi 5/2^-_{\sf 5}\otimes
 \nu 13/2^-_{\sf 1} \otimes \nu 9/2^+_{\sf 4}\}$ are the most favoured in
 $^{271}$Bh ($N=164$), at estimated 0.7 (0.3) MeV excitation energy above
 the rotational g.s. band.

%%%%%%%%%%%%%%%%%%%%%%%%%%%%%%%%%%%%%%%%%%%%%%%%%%%%%%%%%%%%%%%%%%%%%%%%%%%%%%%%%%%%%%%%%%%%%%%%%%%
 {\it Mt\ \ } Among Mt isotopes the best candidate occurs in $^{269}$Mt
 (type B):  $\pi\nu^2 27/2^+\{\pi 11/2^+_{\sf 1}\otimes \nu 9/2^+_{\sf 3}
 \otimes \nu 7/2^+_{\sf 5}\}$, at estimated 0.1 MeV above the rotational g.s.
 band; the same configuration is the lowest one in
 $^{267}$Mt, but already at the excitation energy by 150 keV higher
 than in $^{269}$Mt (see Fig.  \ref{1p2n_109}).
  The second lowest state in $^{269}$Mt:
 $\pi\nu^2 31/2^-\{\pi 11/2^+_{\sf 1}\otimes \nu 9/2^+_{\sf 3} \otimes
 \nu 11/2^-_{\sf 2}\}$ lies more than 200 keV above the lowest one.
 Two configurations, $\pi\nu^2 29/2^- \{\pi 11/2^+_{\sf 1}\otimes \nu 11/2^-_{\sf 2} \otimes \nu 7/2^+_{\sf 5}\}$, $\pi\nu^2 31/2^- \{\pi 11/2^+_{\sf 1}\otimes \nu 11/2^-_{\sf 2} \otimes \nu 9/2^+_{\sf 3}\}$,
 are the lowest ones in $^{265}$Mt. The type-C) candidates in $^{273}$Mt are: $\pi\nu^2 29/2^- \{\pi 11/2^+_{\sf 1}\otimes \nu 13/2^-_{\sf 1}\otimes \nu 5/2^+_{\sf 8}\}$, $\pi\nu^2 33/2^-\{\pi 11/2^+_{\sf 1}\otimes \nu 13/2^-_{\sf 1}\otimes \nu 9/2^+_{\sf 4}\}$, the one with lower $K$ lying 200 keV
 lower. The first one is also the lowest one in the $N=166$ isotone.
 Two lowest
 type-A) configurations are: $\pi\nu^2 25/2^+\{\pi 9/2^-_{\sf 2}\otimes \nu 9/2^-_{\sf 3} \otimes \nu 7/2^+_{\sf 5}\}$, $\pi\nu^2 23/2^+\{\pi 9/2^-_{\sf 2}\otimes \nu 9/2^-_{\sf 3} \otimes \nu 5/2^+_{\sf 7}\}$ in $^{259}$Mt.

%%%%%%%%%%%%%%%%%%%%%%%%%%%%%%%%%%%%%%%%%%%%%%%%%%%%%%%%%%%%%%%%%%%%%%%%%%%%%%%%%%%%%%%%%%%%%%%%%%%
{\it Rg\ \ }  In this isotopic chain there are many near-degenerate low-lying
 configurations. The selection shown in Fig. \ref{1p2n_111} in the left panels
correspond to the lowest in the quasiparticle scheme which may be not lowest
 in the blocked BCS scheme.
The best candidates for isomers occur for $N=160$ (type-B):
$\pi\nu^2 19/2^- \{\pi 3/2^-_{\sf 8}\otimes \nu 7/2^+_{\sf 5} \otimes \nu 9/2^+_{\sf 3}\}$, $\pi\nu^2 25/2^- \{\pi 9/2^-_{\sf 2} \otimes \nu 7/2^+_{\sf 5} \otimes \nu 9/2^+_{\sf 3}\}$, $\pi\nu^2 27/2^+\{\pi 11/2^+_{\sf 1} \otimes \nu 7/2^+_{\sf 5} \otimes \nu 9/2^+_{\sf 3}\}$ in $^{271}$Rg and for $N=164$ (type C): $\pi\nu^2 21/2^+\{\pi 3/2^-_{\sf 8} \otimes \nu 5/2^+_{\sf 8} \otimes \nu 13/2^-_{\sf 1}\}$, $\pi\nu^2 19/2^+ \{\pi 1/2^-_{\sf 11} \otimes \nu 5/2^+_{\sf 8} \otimes \nu 13/2^-_{\sf 1}\}$, $\pi\nu^2 29/2^-\{\pi 11/2^+_{\sf 1}\otimes \nu 5/2^+_{\sf 8} \otimes \nu 13/2^-_{\sf 1}\}$ in $^{275}$Rg, see Fig. \ref{1p2n_111}. The C)-type state with $\pi 3/2^-_{\sf 8}$ has the smallest excitation energy of all in Rg isotopes, 1.07 MeV, at an estimated 0.33 MeV
 above the rotational g.s. band. The excitation energy of the same
 configuration in the $N=166$ isotope is $\sim$ 200 keV higher.
 There are less favorable cases in four lighter isotopes:
 $\pi\nu^2 25/2^+ \{\pi 9/2^-_{\sf 2} \otimes
 \nu 9/2^-_{\sf 3}\otimes \nu 7/2^+_{\sf 4}\}$,
 $\pi\nu^2 17/2^+\{\pi 9/2^-_{\sf 2} \otimes \nu 9/2^-_{\sf 3}\otimes \nu 5/2^+_{\sf 7}\}$ in $^{261}$Rg,
 and:
$\pi\nu^2 29/2^-(27/2^+)\{\pi 11/2^+_{\sf 1}(9/2^-_{\sf 2})\otimes
 \nu 11/2^-_{\sf 2}\otimes \nu 7/2^+_{\sf 5}\}$,
$\pi\nu^2 29/2^+(31/2^-)\{\pi 9/2^-_{\sf 2}(11/2^+_{\sf 1})\otimes
 \nu 11/2^-_{\sf 2}\otimes \nu 9/2^+_{\sf 3}\}$,
$\pi\nu^2 25/2^-(27/2^+)\{\pi 9/2^-_{\sf 2}(11/2^+_{\sf 1})\otimes
 \nu 9/2^+_{\sf 3}\otimes \nu 7/2^+_{\sf 5}\}$
 in $^{265,267,269}$Rg.

\begin{figure}[!htbp]
\centerline{\includegraphics[scale=0.55]{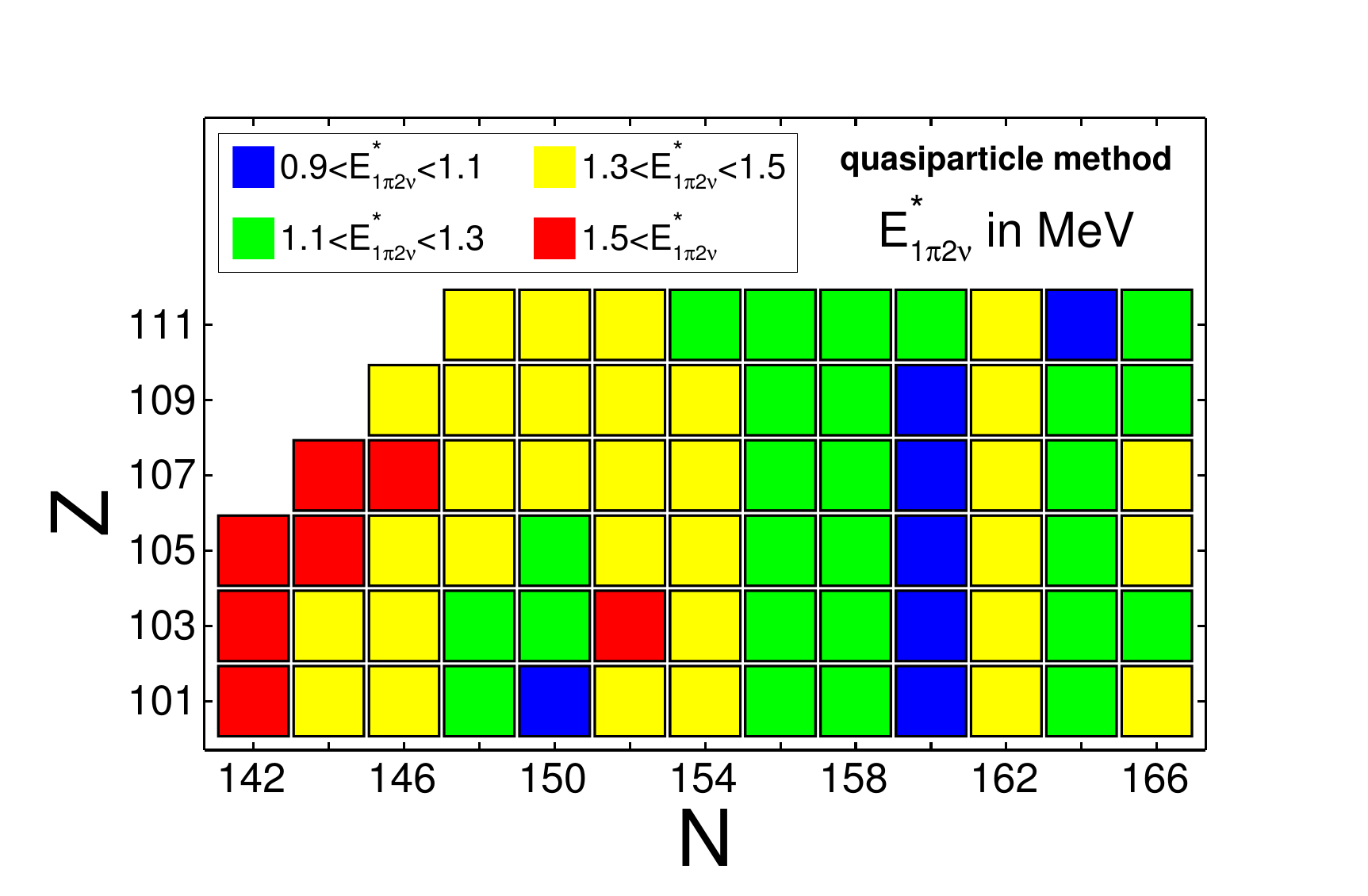}}
\vspace{-5mm}
\caption{{\protect  Map of the lowest $1\pi2\nu$ 3-q.p. excitation energies calculated
 with the quasiparticle method.
 }}
\label{szachnq}
\end{figure}

 Fig. \ref{szachnq} provides a summary of excitation energies for $1\pi2\nu$ q.p.
 configurations from the quasiparticle method.
 The lowest 3-q.p. energies in each isotope obtained from the PNP calculation
 are by 0 - 250 keV larger than from the quasiparticle method (except for
 $N=152,162$ for which the differences are larger), and so are their estimated
 excitation energies above the rotational g.s. band. Hence,
 by the energy criterion suggested by data on known $K$-isomers, the PNP
 calculation also predicts configurations pointed out above as candidates for
 isomers. In particular, the candidates in $N=148, 150$ for Md, Lr, Db, and in
 $N=158, 160, 164$ isotones seem promising.

\subsection{Excitation energies of 3-proton q.p. high-$K$ - candidates for
 isomeric states}

There are fewer 3-proton q.p. configurations than the $1\pi2\nu$ ones due to a smaller density of s.p. proton levels and
lower $\Omega$-values of relevant orbitals. The subshell gap at $Z=108$ in the deformed WS proton spectrum rises energy of
the $3\pi$ excitations in Bh ($Z$=107) isotopes. Even without pairing, their (particle-hole) energies are larger than 1 MeV
(or even larger than 1.5 MeV if one excludes the lightest three and the heaviest isotope).
The blocked BCS results are unreliable as the proton pairing gap vanishes in 3.q.p configurations when one uses pairing strengths of our mass model.
All excitation energies obtained from the
quasiparticle method are greater than 1.4 MeV, i.e. larger than energies of many $1\pi2\nu$ states
(they are listed in Table provided in the supplement material).
The lowest $3\pi$ q.p. excitation energies obtained from the PNP method are smaller than that, in the 0.9 - 1.0 MeV range,
but they occur only for a few configurations. These smallest PNP energies are smaller than $2\Delta$(BCS), where the latter is understood as
the proton pairing energy gap in the g.s. This might suggest a slightly too small proton pairing strength, but, on the other hand, one may expect a weaker pairing in the $3\pi$ q.p. state than in the g.s., so the excitation energies from the quasiparticle method are very likely overestimated.

\begin{figure}[!htbp]
\centerline{\includegraphics[scale=0.8]{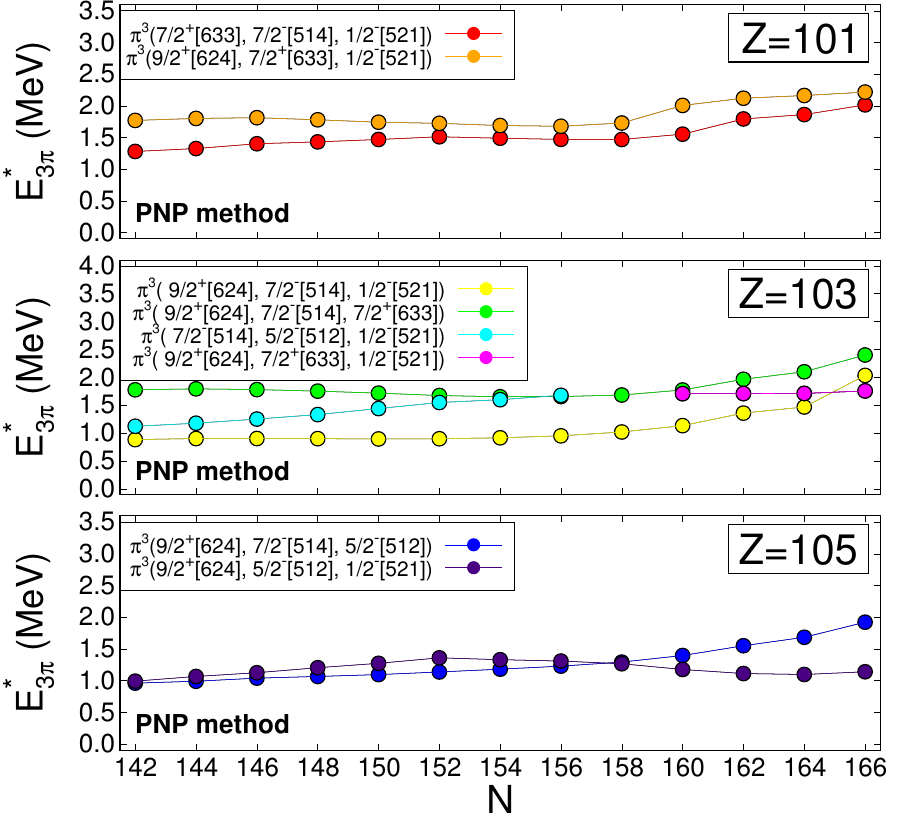}}
\caption{{\protect  Excitation energies of low-lying $3\pi$ configurations
 in Md, Lr and
% ($7/2^-_{\sf 3}\otimes 9/2^+_{\sf 2}\otimes 1/2^-_{\sf 10}$ - blue dots,
% $7/2^-_{\sf 3}\otimes 5/2^-_{\sf 5}\otimes 1/2^-_{\sf 10}$ - red triangles,
% $7/2^+_{\sf 3}\otimes 9/2^+_{\sf 2}\otimes 1/2^-_{\sf 10}$ - green triangles,
% $7/2^-_{\sf 3}\otimes 9/2^+_{\sf 2}\otimes 7/2^-_{\sf 3}$ - black squares),
  Db with the PNP calculation.
}}
\label{3pPNPa}
\end{figure}

\begin{figure}[!htbp]
\centerline{\includegraphics[scale=0.8]{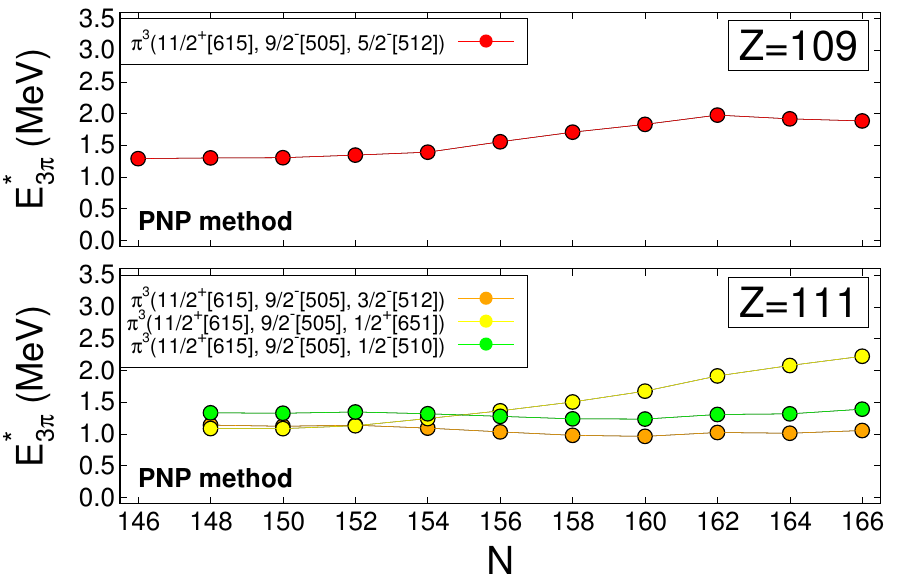}}
\caption{{\protect  Excitation energies of low-lying $3\pi$ configurations in Mt and Rg with the PNP calculation.
}}
\label{3pPNPb}
\end{figure}

\begin{figure}[!htbp]
\centerline{\includegraphics[scale=0.8]{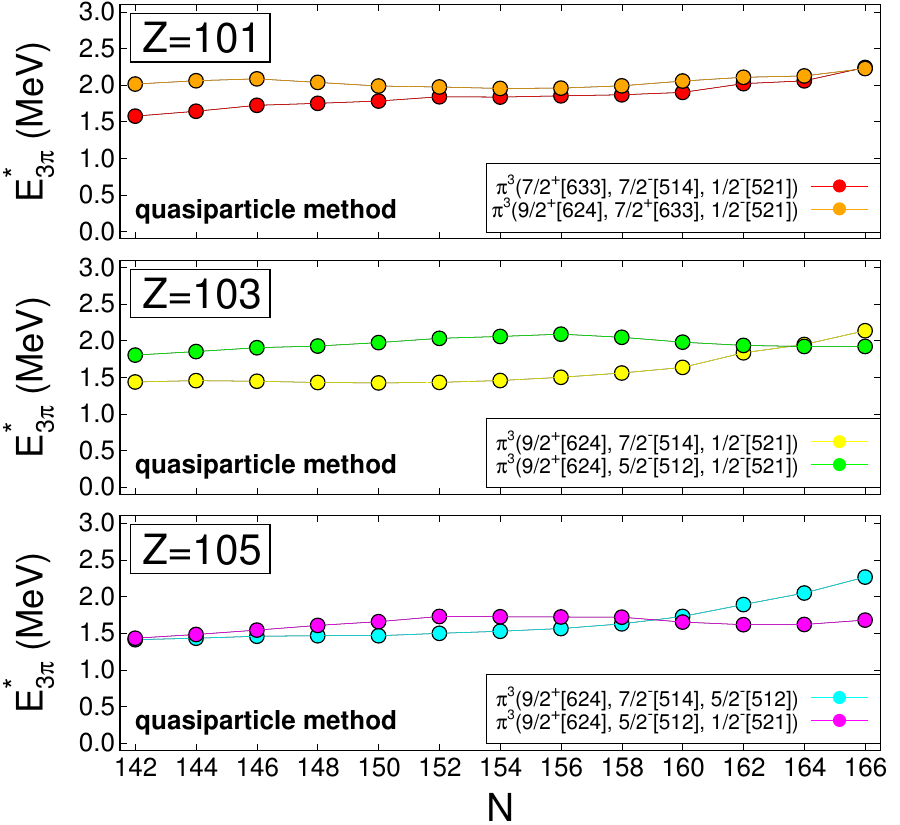}}
%\vspace{-20mm}
\caption{{\protect  Excitation energies of low-lying $3\pi$ configurations
 in Md, Lr and Db with the quasiparticle calculation.
 }}
\label{3pqa}
\end{figure}

\begin{figure}[!htbp]
\centerline{\includegraphics[scale=0.8]{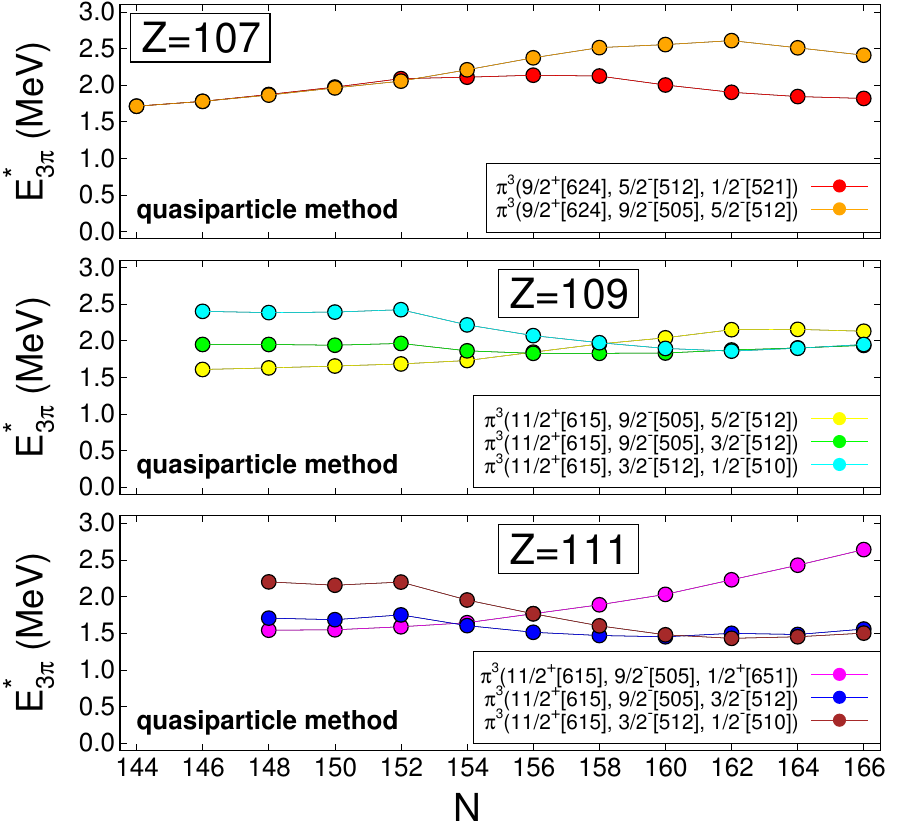}}
%\vspace{-20mm}
\caption{{\protect  Excitation energies of low-lying $3\pi$ configurations
 in Bh, Mt and Rg with the quasiparticle calculation.
 }}
\label{3pqb}
\end{figure}

Energies for favored 3-proton q.p. configurations from the PNP and quasiparticle calculations are shown in Figs. \ref{3pPNPa}, \ref{3pPNPb} and Figs \ref{3pqa}, \ref{3pqb}, respectively. In each of these two methods, the lowest $3\pi$ states were independently selected. As in the case of $1\pi2\nu$ states, the obtained excitation energies $E_{3\pi}^{*}$ depend on the employed pairing version, but the configurations themselves do not.  Below, we discuss the results of  PNP calculations.

%%%%%%%%%%%%%%%%%%%%%%%%%%
%%%%%%%%%%%%%%%%%%%%%%%%%%
{\it Md\ \ } The lowest $3\pi$ q.p. excitation in all studied isotopes is
 $\pi^3 15/2^+\{\pi 7/2^+_{\sf 3}\otimes \pi 7/2^-_{\sf 3}\otimes \pi 1/2^-_{\sf 10}\}$,
 with energy 1.28 - 1.47 MeV, gently rising from $N=142$ to $N=158$; then
 the rise becomes steeper, with energy reaching 2 MeV at $N=166$.
 The second configuration presented in the upper panel in Fig. \ref{3pPNPa},
 $\pi^3 17/2^-\{\pi 7/2^+_{\sf 3}\otimes \pi 9/2^+_{\sf 2}\otimes \pi 1/2^-_{\sf 10}\}$ (orange dots),
 occurs as the second lowest in isotopes from $N=150$ to $N=166$ with
 energies 1.75 - 2.2 MeV. In lighter isotopes, two configurations appear
 below it:
 $\pi^3 11/2^-\{\pi 7/2^-_{\sf 3}\otimes \pi 3/2^-_{\sf 7}\otimes \pi 1/2^-_{\sf 10}\}$
 and in $N=142$ also:
 $\pi^3 19/2^+\{\pi 7/2^+_{\sf 3}\otimes \pi 7/2^-_{\sf 3}\otimes \pi 5/2^-_{\sf 5}\}$.

{\it Lr\ \ } The configuration
 $\pi^3 17/2^+\{\pi 7/2^-_{\sf 3}\otimes \pi 9/2^+_{\sf 2}\otimes \pi 1/2^-_{\sf 10}\}$ (yellow dots in Fig. \ref{3pPNPa})
  is the lowest one in almost all considered isotopes (the only exception is $N=166$). Its excitation energy stays
 nearly constant at  0.9 - 1.0 MeV between $N=142$ and $N=158$; then it rises
 to 2 MeV at $N=166$. If one accepts the experimental assignment of $1/2^-$
 g.s. for $^{253}$Lr ($7/2^-$ is predicted from the WS spectrum), then the
 PNP energy for the state with additional two protons coupled to $K_{2\pi}=8$
  would translate to $\approx$ 0.55 MeV excitation above the g.s. rotational
 band. The configuration
 $\pi^3 13/2^-\{\pi 7/2^-_{\sf 3}\otimes \pi 5/2^-_{\sf 5}\otimes \pi 1/2^-_{\sf 10}\}$
 is the second lowest in $N=142 - 154$. At $N=156, 158$ the
 $\pi^3 23/2^-\{\pi 7/2^+_{\sf 3}\otimes \pi 7/2^-_{\sf 3}\otimes \pi 9/2^+_{\sf 2}\}$,
 configuration becomes the second lowest. The configuration
 $\pi^3 23/2^-\{\pi 9/2^+_{\sf 2}\otimes \pi 7/2^+_{\sf 3}\otimes \pi 1/2^-_{\sf 10}\}$
 (magenta dots in Fig. \ref{3pPNPa}) is the second lowest for
 $N=160 - 166$.

{\it Db  \ } Two configurations shown in Fig. \ref{3pPNPa} are:
 $\pi^3 21/2^+\{\pi 7/2^-_{\sf 3}\otimes \pi 9/2^+_{\sf 2}\otimes \pi 5/2^-_{\sf 5}\}$ and
 $\pi^3 15/2^+\{\pi 5/2^-_{\sf 5}\otimes \pi 9/2^+_{\sf 2}\otimes \pi 1/2^-_{\sf 10}\}$.
 The first one is the lowest one for $N<160$, the second one
  for $N=160-166$. Except for those two and the similar
  configuration
 $\pi^3 17/2^+\{\pi 7/2^-_{\sf 3}\otimes \pi 9/2^+_{\sf 2}\otimes \pi 1/2^-_{\sf 10}\}$ all
 others have considerably larger excitation energies. Estimates of energy
 difference from the g.s. rotational band built on the $\pi 9/2^+_{\sf 2}$
 state are: 0.65 MeV for the first ($K_{2\pi}=6$), and 0.95 MeV for the second
 ($K_{2\pi}=3$) configuration, so the first one seems a better
 candidate for the $K$-isomer.

{\it Mt  \ } In Fig. \ref{3pPNPb} is shown energy of the configuration
 $\pi^3 25/2^+\{\pi 11/2^+_{\sf 1}\otimes \pi 9/2^+_{\sf 2}\otimes \pi 5/2^-_{\sf 5}\}$
 which is the lowest one for $N\le 154$. As energies of all $3\pi$ 3-q.p. states
 rise with $N$, it is probably the best candidate in lighter Mt isotopes.
 For the $\pi 11/2^+$ g.s. which follows from the PNP calculation and
 $K_{2\pi}=7$, one estimates 0.75 MeV excitation above the rotational g.s.
 band for this candidate in $N=156$.

{\it Rg  \ } The configuration
 $\pi^3 23/2^-\{\pi 11/2^+_{\sf 1}\otimes \pi 9/2^+_{\sf 2}\otimes \pi 3/2^-_{\sf 8}\}$ (orange dots in Fig. \ref{3pPNPb}),
 the lowest one for $N=154$-160, seems to be the most interesting candidate.
 For the $\pi 9/2^+_{\sf 2}$ g.s. following from PNP calculations, with
 $K_{2\pi}=7$, one obtaines for $N=160$, $\approx$ 250 keV for estimated
 excitation above the rotational g.s. band.

%%%%%%%%%%%%%%%%%%%%%%%%%%%%%%%%%%%%%%%%%%%%%%%
%%%%%%%%%%%%%%%%%%%%%%%%%%%%%%%%%%%%%%%%%%%%%%%

%%%%%%%%%%%%%%%%%%%%%%%%%%%%%%%%%%%%%%%%%%%%%%%
%%%%%%%%%%%%%%%%%%%%%%%%%%%%%%%%%%%%%%%%%%%%%%%

%%%%%%%%%%%%%%%%%%%%%%%%%%%%%%%%%%%%%%%%%%%%%%%
%%%%%%%%%%%%%%%%%%%%%%%%%%%%%%%%%%%%%%%%%%%%%%%

 The above-mentioned results suggest that low-lying $3\pi$ high-$K$
  configurations appear in Lr, Db, and Rg isotopes. The smallest energies of $3\pi$ configuration occur in Lr. Owing to the s.p. structure,  the favoured 3q.p. configuration in Rg has  a smaller excitation above the g.s. rotational band than the one in Db. The excitation energies $E_{3\pi}^{*}$ in the quasiparticle method are considerably higher (Figs. \ref{3pqa}, \ref{3pqb}), but one should allow
 a  correction for the undiminished g.s. pairing $\Delta$ it uses.
 The $3\pi$ isomers in Mt are uncertain and those in Bh unlikely by the
 energy gap at $Z=108$.

 Regarding our results vs experimental evidence,  the candidates for isomers in $^{249,251}$Md seem to be of the $2\nu 1\pi$ type, in agreement with the interpretation given in
\cite{Goigoux2021}. Configuration similar to
 the one in $^{251}$Md could be isomeric in $^{253}$Lr. On the other hand, a similar
 two-neutron component of the $^{255}$Lr isomer,
 suggested in \cite{Berkeley}, is not supported by
  our calculation as the $2\nu 1\pi$ excitations
  in the $N=152$ isotones are disfavoured, and the
  $3\pi$ configuration looks more favoured, cf Fig. \ref{3pPNPa}, \ref{3pqa}. The same conclusion follows from our results for $^{257}$Db.

\subsection{Possible $\alpha$-decay hindrance of high-$K$ isomers}

 A usual $K$-isomer is recognized by its prolonged $EM$ half-life $T_{\gamma}$. SH nuclei are radioactive and known odd-even SH isotopes undergo  mainly $\alpha$-decay (as their fission is usually hindered, see e.g. review \cite{Hessberger-f}). Thus, there is half-life $T_{\alpha}(gs)$ for the g.s. and the partial half-lives $T_{\alpha}$ and
$T_{\gamma}$ for the isomer. When $T_{\alpha}$ and $T_{\gamma}$ are of distinctively different magnitude, the total half-life of the isomer $T_{\alpha}T_{\gamma}/(T_{\alpha}+T_{\gamma})$ equals the smaller of the two in good approximation.

 The most interesting situation would be a $K$-isomer living longer than the g.s., which requires:  $T_{\alpha}(gs)<min(T_{\alpha}, T_{\gamma})$. There are two possibilities: $T_{\alpha}< T_{\gamma}$ or $T_{\alpha}>T_{\gamma}$. In the first case, isomeric $\gamma$-decay may be more difficult to observe, while two distinct $\alpha$ half-lives $T_{\alpha}(gs)$ and $T_{\alpha}$ can be detected with generally different $Q_{\alpha}$ values.  The  effective half-life of the g.s. $\alpha$-decay after the isomer $\gamma$ deexcitation,    $T_{\gamma}+T_{\alpha}(gs)$, will be statistically suppressed. In the second case, the $\alpha$-decay of the isomer may be difficult to observe, while two $\alpha$ activities will be detected with half-lives  $T_{\alpha}(gs)$ and $T_{\gamma}+T_{\alpha}(gs)$ and common $Q_{\alpha}$; coincidences with gamma/converted electron detectors can  help to  infer something about the isomer $\gamma$-decay
 scheme and excitation energy $E^*(K)$.
  When the three considered half-lives are of the similar order, half-lives of three alpha activities and some $\gamma$ transitions can be detected. When more
isomers are present or multiple $Q_{\alpha}$ values occur, the experimental analysis may become complex.

 As stated in the Introduction, we have no way of predicting $T_{\gamma}$. What can be said about
 the condition $T_{\alpha}(gs)<T_{\alpha}$?

 One possibility, discussed in \cite{Jachimowicz2018}, is that the excitation of an isomeric configuration in the $\alpha$-daughter $E_d^*$ is substantially higher than that of the isomer in the parent nucleus $E_p^*$. As understood from $\alpha$-decay systematics, the configuration-changing $\alpha$-decays, between states with different $K$ and/or parity, are usually hindered (see e.g. \cite{Hessberger-a}, p 40). If the configuration-preserving decay is suppressed by a reduced $Q_{\alpha}=Q_{\alpha}(gs)-\Delta Q_{\alpha}$, where $\Delta Q_{\alpha}=(E_d^*-E_p^*)$ - as in the situation above - the decay may proceed to different lower-lying  configurations. A detailed balance of the corresponding decay-rates - a gain due to a greater $Q_{\alpha}$ (a smaller $Q_{\alpha}$-hindrance) vs a loss due to a larger configuration-hindrance - is difficult to evaluate. However, one may look for cases of reduced $Q_{\alpha}$-value in configuration-preserving isomer decays as
indicative of possible enhanced isomer stability. We note that phenomenological formulas giving $\alpha$ half-live $T_{\alpha}$ as a
function of $Q_{\alpha}$ apparently contain some part of this effect so that, {\it at the same} $Q_{\alpha}$, they predict {\it less
probable} g.s.$\rightarrow$g.s. transitions for odd-$A$ and odd-odd
nuclei than for the neighboring even-even ones, see e.g. \cite{Royer22}.

The reported example of an isotope with $\alpha$-decaying $K$-isomer living longer than the g.s. is $^{270}$Ds.  The $\alpha$-decay half-life of the 6 ms isomer, with the proposed spin between 8 and 10, is very likely shorter than its $\gamma$-decay half-life - see e.g. the estimate in \cite{Batar2022}, but longer than the approximately 0.1 ms half-life  of the $\alpha$-decaying ground state \cite{Hofmann2001, AckermannGSI}.
One could suspect that the high-$K$ state with a similar structure exists in the daughter nucleus $^{266}$Hs; indeed, there is also evidence for a $K$-isomeric state in $^{266}$Hs  \cite{AckermannGSI}. The  $Q_{\alpha}$-hindrance factor effect on the half-life of the $^{270}$Dm isomer
was estimated within the MM Woods-Saxon model in \cite{Jachimowicz2018}. It turns out that its logarithm is three times larger than the experimental value, so the
effect seems to exist, but is considerably weaker than the pure $Q_{\alpha}$-hindrance.

We looked for $Q_{\alpha}$-hindrance  among the lowest - lying $1\pi2\nu$ high-$K$ configurations
in the present calculations. The $\Delta Q_{\alpha}$ for structure-preserving
$\alpha$-decays show maxima at $N=154$ and $N=164$, as dictated by the subshell
 gaps in the WS s.p. spectrum and seen in Fig. \ref{1p2n_101}-\ref{1p2n_111}.
 The former is bigger in Md, Lr, the latter in Db-Rg, and especially in Mt isotopes. The $\Delta Q_{\alpha}$ obtained in the BCS with blocking are larger than in the quasiparticle method.  The largest $\Delta Q_{\alpha}$ occurs in $^{273}$Mt: 3.27 MeV in the blocked BCS, 1.62 MeV in the quasiparticle method, and 2.60 MeV in the PNP. Below,  we describe this extreme case using PNP energies for the purpose of illustration .

The $Q_{\alpha}$-hindrance for the decay preserving the lowest-lying $\pi\nu^2$ configuration in $^{273}$Mt,  $29/2^- \{\pi 11/2^+_{\sf 1}\otimes \nu 13/2^-_{\sf 1}\otimes \nu 5/2^+_{\sf 8}\}$, would imply a prolongation of $T_{\alpha}$ by $\approx$8 orders of magnitude
 (roughly, 3 orders per 1 MeV $\Delta Q_{\alpha}$).
However, there are many lower-lying possible final $\pi\nu^2$ configurations in $^{269}$Bh; among them one
with $\pi 5/2^-_{\sf 5}$ instead of $\pi 11/2^+_{\sf 1}$, $\Delta K=-3$,
 $\Delta Q_{\alpha}\approx 1.94$ MeV,
 and one with $\nu 7/2^+_{\sf 5}$ instead of $\nu 5/2^+_{\sf 8}$,
 $\Delta K=1$ and $\Delta Q_{\alpha}\approx 1.43$ MeV, which differ from the  initial state {\it by only one quasiparticle}. There is also a state $29/2^- \{\pi 9/2^+_{\sf 2}\otimes \nu 13/2^-_{\sf 1}\otimes \nu 7/2^+_{\sf 5}\}$, with the same $K$ and parity, differing by two quasiparticles, for which
 $\Delta Q_{\alpha}=1.08$ MeV. The presence of many configurations for which
 the structural hindrance is (at least partly) cancelled by a smaller
 $Q_{\alpha}$-hindrance suggests that, if the  configuration in $^{273}$Mt
 really turns out to be isomeric, its $\alpha$-decay hindrance will be smaller
 than the $Q_{\alpha}$- hindrance. Nevertheless, as long as the predicted
 subhell gaps are realistic, and all 3qp high-$K$ configurations in $^{269}$Bh
 lie higher  than in $^{273}$Mt, cf Fig.  \ref{1p2n_107},\ref{1p2n_109},
 some $\alpha$-decay hindrance of  the considered configuration
 should be expected.

% The most prominent cases of smaller $Q_{\alpha}$ values occur according to
%  subshell gaps in the WS s.p. spectrum .....

%At the end it is worth mentioning the predictions for
%the recently examined $^{255}$ Lr by Ketelhut, et al., \cite{Ketelhut2009},
%or Jeppesen, et al., \cite{Jeppesen2009},
%while neighbouring nuclei $^{253,257}$ Lr are the natural next heavy odd systems to be studied.

%FOR 253 Lr
%103 150 -7/2 [5 1 4] -9/2 [7 3 4] 7/2 [6 2 4] E*=0.8 MeV

%N 255Lr OUR CANDITATE (LOW-LYING CONFIGURATION) FROM BLOCKING SCENARIO IS:
%7/2- [514] (proton), 9/2- [734] (1 neutron), 11/2- [725] (2 neutron) WITH E*=1.3 MeV .

%FOR 257 Lr
%103 154 -7/2 [5 1 4] -11/2 [7 2 5] 7/2 [6 1 3] E*=1.1 MeV

\section{CONCLUSIONS}

Considering both $1\pi2\nu$ and $3\pi$ 3-q.p. excitations within the MM Woods-Saxon model we have
found the candidates for high-$K$ isomers in odd-even Md - Rg isotopes. Using various treatments of pairing we showed that, although they
change calculated excitation energies, the favored configurations remain the same. Thus, the presented results may be treated as specific to the
used MM model and determined primarily by the s.p. spectrum of the WS potential. For the same reason, predicted isomer energies should be
understood as approximate.

The characteristic features of the results are the numbers $Z$ and $N$ of isotopes in which the high-$K$ states have the
lowest excitation energies and the structure of those favored configurations. Particularly favoured  neutron numbers for $1\pi2\nu$
states are: $N=148, 150, 158, 160, 164$; the lowest configurations occur in Md for $N=150$, in Bh for $N=160$, and in Rg for $N=164$.
A few low-lying $3\pi$ configurations occur in Lr, Db, and Rg isotopes; those in Lr and Rg seem more promising due to an estimated smaller excitation over the g.s. rotational band.
Our results are consistent with the interpretation of experimental data
 for $K$-isomers in $N=148,150$ isotones of Md and Lr. On the other hand, they suggest  $3\pi$ rather than $1\pi 2\nu$ structure in $N=152$ isotones, as $N=152, 162$ are disfavoured for $K$-isomers owing to neutron subshell gaps, as is $Z=108$ for protons. Certainly, more experimental data are needed to check theoretical predictions
  and the related s.p. level scheme.

 A possible hindrance of a $K$-isomer $\alpha$-decay could make it more
long-lived than the g.s. The $Q_{\alpha}$-hindrance can be a reason for such a situation if the structural hindrance is sufficiently strong. The strongest $Q_{\alpha}$-hindrance follows from our model for $N=164$ and
 for heavier isotopes of Mt, with the maximum $\Delta Q_{\alpha}$ for $^{273}$Mt.

 %The structure of the candidate configurations is displayed in Tables in the Supplement Material.

Clearly, the predictions presented here, based on the extrapolation only partially rooted in experimentally established facts, should be
subjected to experimental tests. We hope that they will appear useful in the ongoing research on superheavy nuclei.

\section*{ACKNOWLEDGEMENTS}

M.~K. was co-financed by the International Research Project COPIGAL.

\appendix

\section{Particle-number-projected BCS calculations}

 We specify here procedures used for calculating energy for a state of the
 BCS-form projected onto a good particle number (separately for neutrons
 and protons). For PNP to have an effect in the weak pairing limit, one
 admits parameters $\lambda$ and
 $\Delta$ unconstrained by the BCS equations. As the projection method
 itself is well known, see e.g. \cite{RingSchuck}, the present account is rather brief.

 The squared norm, $\langle\Psi\mid{\hat P}_N\mid\Psi\rangle$, of the
 $N=2n$-particle component of the BCS wave function
 $\mid\Psi\rangle=\prod_{\nu>0}
 (u_{\nu}+v_{\nu}a^{\dagger}_{\nu}a^{\dagger}_{\bar \nu})\mid vac\rangle$,
 where $\mid vac\rangle$ is the physical vacuum (no particles) and the label $\nu$
 enumerates states with positive $\Omega$, is
 equal to the term by $\zeta^{n}$ in the expression: $\prod_{\nu>0}
 (u^2_{\nu}+\zeta v^2_{\nu})$, which we call $P_n$. Denoting $P^{\mu}_{n-1}$
 the squared norm of the $(N-2)$-particle component of $\mid\Psi\rangle$ with the
 {\it omitted} factor
$(u_{\mu}+v_{\mu}a^{\dagger}_{\mu}a^{\dagger}_{\bar \mu})$,
 and $P^{\mu \nu}_{n-1}$ - the squared norm of the $(N-2)$-particle component
 of $\mid\Psi\rangle$ with the {\it omitted} factors
$(u_{\mu}+v_{\mu}a^{\dagger}_{\mu}a^{\dagger}_{\bar \mu})
 (u_{\nu}+v_{\nu}a^{\dagger}_{\nu}a^{\dagger}_{\bar \nu})$, one can write
 the energy of the $N$-particle projected BCS state as:
 \begin{equation}
 \label{Erzut}
  E_N=\frac{\sum_{\mu>0}(2\epsilon_{\mu}-G)v^2_{\mu}P^{\mu}_{n-1} -
     G\sum_{\mu>0\ne\nu>0}u_{\mu}v_{\mu}u_{\nu}v_{\nu}P^{\mu \nu}_{n-1}}
     {P_n}  .
 \end{equation}
 The first way to calculate $E_N$ utilizes a representation of $\zeta$ as
 the $(n+1)$-dimensional Jordan block with zero on the diagonal, that means
 a matrix with elements of the first upper diagonal equal to 1 and all
 others equal zero. A simple matrix multiplication leads to $P_n$ being
 the entry $n+1,n+1$ of the matrix $\prod_{\nu>0} (u^2_{\nu}+\zeta v^2_{\nu})$.
  After calculating all $P^{\mu}_{n-1}$ in the same way and using the identity:
  $(u^2_{\nu}v^2_{\mu}-u^2_{\mu}v^2_{\nu})P^{\mu \nu}_{n-1}=
 v^2_{\mu}P^{\mu}_{n-1}-v^2_{\nu}P^{\nu}_{n-1}$, one can calculate
 (\ref{Erzut}).

 In the second method, one uses successive substitutions $\zeta=e^{i\varphi_k}$,
 with  $\varphi_k=2 k \pi/M$, $k=$0, 1, ...,$M-1$, and sums over $k$ the
  expressions:
 \begin{equation}
 e^{-in\varphi_k}\prod_{\beta>0}(u^2_{\beta}+e^{i\varphi_k}v^2_{\beta})
 \left(\sum_{\mu>0}\frac{(2\epsilon_{\mu}-G)v^2_{\mu}e^{i\varphi_k}}
 {u^2_{\mu}+e^{i\varphi_k}v^2_{\mu}} -  \\ G\sum_{\mu>0\ne\nu>0}
 \frac{u_{\mu}v_{\mu}u_{\nu}v_{\nu}e^{i\varphi_k}}
{(u^2_{\mu}+e^{i\varphi_k}v^2_{\mu})
 (u^2_{\nu}+e^{i\varphi_k}v^2_{\nu})}\right) .
 \end{equation}
 The formula: $\sum_{k=0}^{M-1} e^{i\varphi_k} = (1-e^{iM\varphi_k})/
 (1-e^{i\varphi_k})$ guarantees that so calculated $E_N$ contains only
 contributions of $\mid\Psi\rangle$ components with particle numbers: $2n$,
 $2n\pm M$, $2n\pm 2M$, etc. In practical calculations, the value of $M$
 like 10 or 15 gives a sufficient accuracy.

 We applied both methods described above and checked that they give the same
 results for $E_N$.

 In order to reduce a search for the optimal projected BCS state to a one-parameter minimization, we eliminate the dependence of projected energies $E_N$ on $\lambda$
 by imposing the condition for the expected number of particles
 on each BCS-type function $\Psi(\lambda,\Delta)$:
 $\langle \Psi_{BCS}(\lambda,\Delta)\mid{\hat N}\mid
 \Psi_{BCS}(\lambda,\Delta)\rangle=N$, where $N$ is a desired value.
 The so obtained energy $E_N(\Delta)$ has a unique minimum as a function of
 $\Delta$. This minimal value is taken as the energy after projection and
  minimized over deformations.

%%%%%%%%%%%%%%%%%%%%%%%%%%%%%%%%%%%%%
%\begin{figure}[!htbp]
%\centerline{\includegraphics[scale=0.8]{EGS_blk_vs_PNP.pdf}}
%\caption{{\protect  Ground-state energy difference between %the blocking and
%PNP calculation for six $Z$-odd, $N$-even isotopic chains: %Md - Rg. %[suplement material].
%}}
%\label{engsrozn}
%\end{figure}
%%%%%%%%%%%%%%%%%%%%%%%%%%%%%%%%%%%%%

%%%%%%%%%%%%%%%%%%%%%%%%%%%%%%%%%%%%%
\begin{figure}[!htbp]
\centerline{\includegraphics[scale=0.8]{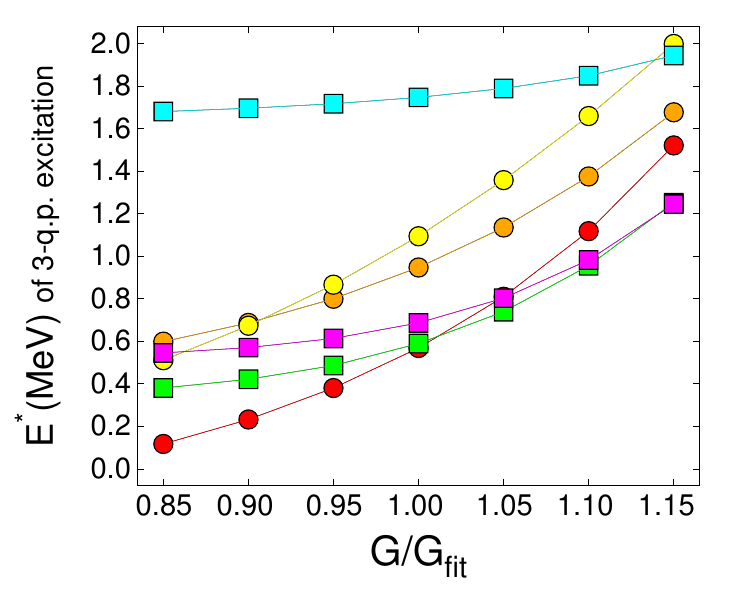}}
\caption{{\protect Energy of the 3-q.p. excitation ($E_{1\pi2\nu}^{*}$ or $E_{3\pi}^{*}$ in MeV) from the
 PNP method vs pairing strength (in units of the value of the fit
 $G/G_{fit}$) at the fixed, near g.s. shape, for $1\pi2\nu$ states in Mt isotopes
 (dots):
 $\pi 9/2^-_{\sf 2}\otimes \nu 7/2^+_{\sf 4}\otimes\nu 5/2^+_{\sf 7}$ in $^{255}$Mt (yellow),
 $\pi 11/2^+_{\sf 1}\otimes \nu 9/2^+_{\sf 3}\otimes\nu 7/2^+_{\sf 5}$
 in $^{267}$Mt (orange) and $^{269}$Mt (red), and for $3\pi$ states (squares):
 $\pi 7/2^-_{\sf 3}\otimes \pi 9/2^+_{\sf 2}\otimes\pi 1/2^-_{\sf 10}$ in $^{259}$Lr (green),
 $\pi 11/2^+_{\sf 1}\otimes \pi 9/2^-_{\sf 2}\otimes\pi 5/2^-_{\sf 5}$ in $^{265}$Mt (cyan), and
 $\pi 9/2^+_{\sf 2}\otimes \pi 11/2^+_{\sf 1}\otimes\pi 3/2^-_{\sf 8}$ in $^{271}$Rg (magenta).
 }}
\label{rzutG}
\end{figure}
%%%%%%%%%%%%%%%%%%%%%%%%%%%%%%%%%%%%%

  The excitation energy of a state with extra two blocked particles relative to
  the odd-even g.s. includes the effect of deformation change. However, at a fixed
deformation, it is a function of the pairing strength $G$, the neutron one for a $1\pi2\nu$, and the proton one for a $3\pi$ q.p. configuration. This function is shown in Fig. \ref{rzutG} for
  three $1\pi2\nu$ and three $3\pi$ configurations; the strength $G$ is
  expressed in units of $G_{fit}$, where $G_{fit}$ is the value from our mass
  model \cite{Jachimowicz2014,Jachimowicz2021} for a given nucleus. The rise in the excitation energy with $G$ is
  the result of a decrease in energy of both the g.s. and 3-q.p. state,
  with the former being steeper than the latter.


\begin{thebibliography}{99}

\bibitem{Walker1999}
P. Walker and G. Dracoulis, Nature \textbf{399}, 35 (1999).

\bibitem{Batar}
 J. Khuyagbaatar, EPJ Web Conf. \textbf{163}, 00030 (2017).

\bibitem{Lopez}
 A. Lopez-Martens \textit{et al.}, Eur. Phys. J. A \textbf{58}, 134 (2022).

\bibitem{Herzberg}
 R. D. Herzberg and P. Greenlees, Prog. Part. Nucl. Phys. \textbf{61}, 674 (2008).

\bibitem{kon2}
 F. G. Kondev \textit{et al.}, At. Data Nucl. Data Tables \textbf{103-104},  (2015).

\bibitem{Walker2016}
 P. M. Walker and F. R. Xu, Phys. Scr. \textbf{91}, 013010 (2016).

\bibitem{Ghiorso1973}
 A. Ghiorso \textit{et al.}, Phys. Rev. C \textbf{7}, 2032 (1973).

\bibitem{Tandel2006}
 S. K. Tandel \textit{et al.}, Phys. Rev. Lett. \textbf{97}, 082502 (2006).

\bibitem{Herzberg2006}
 R. D. Herzberg \textit{et al.}, Phys. Rev. Lett. \textbf{97}, 082502 (2006).

\bibitem{HerzbergNAT2006}
 R.-D. Herzberg \textit{et al.}, Nature \textbf{442}, 996 (2006).

\bibitem{Hessberger2010}
 F. P. Heßberger \textit{et al.}, Eur. Phys. J. A \textbf{43}, 55 (2010).

\bibitem{Clark2010}
 R. M. Clark \textit{et al.}, Phys. Lett. B \textbf{690}, 19 (2010).

\bibitem{Theisen2015}
 Ch. Theisen \textit{et al.}, Nucl. Phys. A \textbf{944}, 333 (2015).

\bibitem{Robinson2008}
 A. P. Robinson \textit{et al.}, Phys. Rev. C \textbf{78}, 034308 (2008).

\bibitem{Sulignano2012}
 B. Sulignano \textit{et al.}, Phys. Rev. C \textbf{86}, 044318 (2012).

\bibitem{Kallunkathariyil2020}
 J. Kallunkathariyil \textit{et al.}, Phys. Rev. C \textbf{101}, 011301(R) (2020).

\bibitem{Peterson2006}
 D. Peterson \textit{et al.}, Phys. Rev. C \textbf{74}, 014316 (2006).

\bibitem{David2015}
 H. M. David \textit{et al.}, Phys. Rev. Lett. \textbf{115}, 132502 (2015).

\bibitem{Hauschild2008}
 K. Hauschild \textit{et al.}, Phys. Rev. C \textbf{78}, 021302(R) (2008).

\bibitem{GSI}
 S. Antalic \textit{et al.}, Eur. Phys. J. A \textbf{38}, 219 (2008).

\bibitem{Berkeley}
 H.B. Jeppesen \textit{et al.}, Phys. Rev. C \textbf{80}, 034324 (2009).

\bibitem{Goigoux2021}
 T. Goigoux \textit{et al.}, Eur. Phys. J. A \textbf{57}, 321 (2021).

\bibitem{Chatillon}
 A. Chatillon \textit{et al.}, Eur. Phys. J. \textbf{30}, 397 (2006).

\bibitem{Asai2015}
 M. Asai, F.P. Heßberger, and A. Lopez-Martens, Nucl. Phys. A \textbf{944} (2015).

\bibitem{Theissen2020}
 R. Briselet \textit{et al.}, Phys. Rev. C \textbf{102}, 014307 (2020).

\bibitem{Lopez2}
 A. Lopez-Martens and K. Hauschild, Eur. Phys. J. A \textbf{58}(7), 134 (2022).

\bibitem{Hessberger2001}
 F. Heßberger \textit{et al.}, Eur. Phys. J. A \textbf{12}, 57 (2001).

\bibitem{Ackermann2015}
 D. Ackermann, Nucl. Phys A \textbf{944} (2015).

\bibitem{Dracoulis2016}
 G. D. Dracoulis, P. M. Walker, and F. G. Kondev, Rep. Prog. Phys. \textbf{79}, 076301 (2016).

\bibitem{Walker2020}
 P. Walker and Z. Podolyák, Phys. Scr. \textbf{95}, 044004 (2020).

\bibitem{Greenlees2008}
 P. T. Greenlees \textit{et al.}, Phys. Rev. C \textbf{78}, 021303(R) (2008).

\bibitem{Liu2014}
 H. L. Liu, P. M. Walker, and F. R. Xu, Phys. Rev. C \textbf{89}, 044304 (2014).

\bibitem{Herzberg2001}
 R.-D. Herzberg \textit{et al.}, Phys. Rev. C \textbf{65}, 014303 (2001).

\bibitem{Ketelhut2009}
 S. Ketelhut \textit{et al.}, Phys. Rev. Lett. \textbf{102}, 212501 (2009).

\bibitem{Eeckhaudt2005}
 S. Eeckhaudt \textit{et al.}, Eur. Phys. J. A \textbf{26}, 227 (2005).

\bibitem{Greenlees2012}
 P. T. Greenlees \textit{et al.}, Phys. Rev. Lett. \textbf{109}, 012501 (2012).

\bibitem{Cwiok1987}
S.~\'Cwiok, J.~Dudek, W.~Nazarewicz, J.~Skalski, and T.~Werner, Comput. Phys. Commun. \textbf{46}, 379 (1987).

\bibitem{Krappe1979}
H. J. Krappe, J. R. Nix, and A. J. Sierk, Phys. Rev. C \textbf{20}, 992 (1979).


\bibitem{Muntian2001}
I. Muntian, Z. Patyk, and A. Sobiczewski,
\textit{Acta Phys. Pol. B} \textbf{32}, 691 (2001).

\bibitem{Kowal2010}
M. Kowal, P. Jachimowicz, and A. Sobiczewski,
\textit{Phys. Rev. C} \textbf{82}, 014303 (2010).

\bibitem{Jachimowicz2014}
P. Jachimowicz, M. Kowal, and J. Skalski,
\textit{Phys. Rev. C} \textbf{89}, 024304 (2014).

\bibitem{Jachimowicz2012-20}
P. Jachimowicz, M. Kowal, and J. Skalski,
\textit{Phys. Rev. C} \textbf{85}, 034305 (2012);
\textit{Phys. Rev. C} \textbf{101}, 014311 (2020).

\bibitem{Jachimowicz2017-2}
P. Jachimowicz, M. Kowal, and J. Skalski,
\textit{Phys. Rev. C} \textbf{95}, 014303 (2017).

\bibitem{Jachimowicz2021}
P. Jachimowicz, M. Kowal, and J. Skalski,
\textit{Atomic Data and Nuclear Data Tables} \textbf{138}, 101393 (2021).

\bibitem{Stefan}
S. Ćwiok and S. Hofmann,
\textit{Nucl. Phys. A} \textbf{573}, 356 (1994).

\bibitem{ParSob}
A. Parkhomenko and A. Sobiczewski,
\textit{Acta Phys. Pol. B} \textbf{35}, 2447 (2004);
\textit{Acta Phys. Pol. B} \textbf{36}, 3115 (2005).

\bibitem{Patyk19911}
Z. Patyk and A. Sobiczewski,
\textit{Nucl. Phys. A} \textbf{533}, 132 (1991).

\bibitem{Patyk19912}
Z. Patyk and A. Sobiczewski,
\textit{Phys. Lett. B} \textbf{256}, 307 (1991).

\bibitem{Liu2011}
H. L. Liu, F. R. Xu, P. M. Walker, and C. A. Bertulani,
\textit{Phys. Rev. C} \textbf{83}, 011303(R) (2011).

\bibitem{Minkov2022}
N. Minkov, L. Bonneau, P. Quentin, J. Bartel, H. Molique, and D. Ivanova,
\textit{Phys. Rev. C} \textbf{105}, 044329 (2020).

\bibitem{momJ}
A. Sobiczewski, I. Muntian, and Z. Patyk,
\textit{Phys. Rev. C} \textbf{63}, 034306 (2001).

\bibitem{Hessberger-f}
F. P. Hessberger,
\textit{Eur. Phys. J. A} \textbf{53}, 75 (2017).

\bibitem{Jachimowicz2018}
P. Jachimowicz, M. Kowal, and J. Skalski,
\textit{Phys. Rev. C} \textbf{98}, 014320 (2018).

\bibitem{Hessberger-a}
F. P. Hessberger,
arXiv:2102.08793 (2021).

\bibitem{Royer22}
G. Royer, Q. Ferrier, and M. Pineau,
\textit{Nucl. Phys.} \textbf{1021}, 122427 (2022).

\bibitem{Batar2022}
J. Khuyagbaatar,
\textit{Eur. Phys. J. A} \textbf{58}, 243 (2022).

\bibitem{Hofmann2001}
S. Hofmann, F. P. Hessberger, et al.,
\textit{Eur. Phys. J. A} \textbf{10}, 5 (2001).

\bibitem{AckermannGSI}
D. Ackermann et al.,
\textit{GSI Sci. Rep. 2011}, 208 (2012).

\bibitem{RingSchuck}
P. Ring, and P. Schuck,
"The Nuclear Many-Body Problem", Springer-Verlag (New York,1980).


\bibitem{Pyatov1964}
N. I. Pyatov and A. S. Chernyehev,
Izv, ANSSR, seria, Fiz. \textbf{28}, 1173 (1964).

\bibitem{Jain1992}
K. Jain and A. K. Jain,
\textit{Phys. Rev. C} \textbf{3013}, (1992).


\end{thebibliography}
\end{document}